%% file: preprint.tex
\documentclass[useAMS,usenatbib]{mn2e}
\usepackage{graphicx}
\usepackage{amssymb}
\usepackage{wasysym}
\usepackage{pifont}
\usepackage[usenames]{color}
\usepackage{colortbl}
\usepackage{xcolor}

\input{aas_macros}

\def\kms{{\rm km}~{\rm s}^{-1}}
\definecolor{Orange1}{rgb}{1.0,0.9,0.7}
\definecolor{Orange2}{rgb}{1.0,0.95,0.85}
\definecolor{Orange3}{rgb}{0.5,0.15,0.00}

\title[Dark halo and MBH scaling relations]{
Dark halo microphysics and massive black hole scaling relations in galaxies
}
\author[Saxton, Soria \& Wu]{Curtis J. Saxton$^{1,2}$\thanks{E-mail:
saxton@physics.technion.ac.il (CJS);
roberto.soria@icrar.org (RS); 
kinwah.wu@ucl.ac.uk (KW)
}, 
Roberto Soria$^{3}$,
Kinwah Wu$^{1}$
\\
$^{1}$Mullard Space Science Laboratory, University College London,
Holmbury St Mary, Surrey RH5 6NT, UK\\
$^{2}$Physics Department, Technion - Israel Institute of Technology, Haifa 32000, Israel\\
$^{3}$International Centre for Radio Astronomy Research, Curtin University, GPO Box U1987, Perth, WA 6845, Australia
}

\begin{document}

\date{Accepted 2014 September 22.  Received 2014 September 22; in original form 2014 March 23}

\pagerange{\pageref{firstpage}--\pageref{lastpage}} \pubyear{2014}

\maketitle

\label{firstpage}

\begin{abstract} 
We investigate the black hole (BH) scaling relation in galaxies
   using a model in which the galaxy halo and central BH
   are a self-gravitating sphere of dark matter (DM)
   with an isotropic, adiabatic equation of state.
The equipotential where the escape velocity
   approaches the speed of light
   defines the horizon of the BH.
We find that the BH mass ($m_\bullet$)
   depends on the DM entropy,
   when the effective thermal degrees of freedom ($F$)
   are specified.
Relations between BH
   and galaxy properties arise naturally,
   with the BH mass and DM velocity dispersion
   following $m_\bullet\propto\sigma^{F/2}$
   (for global mean density set by external cosmogony).
Imposing observationally derived constraints on $F$
   provides insight into the microphysics of DM.
Given
   that DM velocities and stellar velocities are comparable,
   the empirical correlation between $m_\bullet$
   and stellar velocity dispersions $\sigma_\bigstar$
   implies that $7\la{F}<10$.
A link between $m_\bullet$ and globular cluster properties
   also arises because the halo potential
   binds the globular cluster swarm at large radii.
Interestingly, for $F>6$
   the dense dark envelope surrounding the BH
   approaches the mean density of the BH itself,
   while the outer halo can show a nearly uniform kpc-scale core
   resembling those observed in galaxies.
\end{abstract}

\begin{keywords}
	black holes
	---
	dark matter
	---
	galaxies: haloes
	---
	galaxies: kinematics and dynamics
	---
	galaxies: structure
	---
	globular clusters: general
\end{keywords}


\section{Introduction}
\label{s.introduction}

Over the last two decades,
   empirical correlations between different galaxy components
   --- nuclear supermassive black hole (SMBH),
   stellar bulge and disc,
   dark matter (DM) halo
   --- have shaped our understanding of galaxy structure evolution
   and of SMBH/galaxy co-evolution
	\citep[][for a review]{kormendy2013}.
The most significant correlations are
   the one observed between SMBH masses
   ($m_\bullet$)
   and velocity dispersions
   ($\sigma$)
   of their host stellar bulges or spheroids
\citep[$m_\bullet$--$\sigma$ relation:][%
	]{ferrarese2000,gebhardt2000,tremaine2002,graham2011,xiao2011};
   and the one between SMBH masses and bulge masses
	\citep{magorrian1998,laor2001,haering2004,graham2013,scott2013}.
The kinetic or potential energy of the bulge
   also correlates with $m_\bullet$
	\citep{feoli2005,feoli2007,aller2007,hopkins2007b,
		feoli2009,mancini2012,benedetto2013},
   as does the momentum-like quantity $M_\bigstar\sigma/c$
   where $M_\bigstar$ is the bulge stellar mass
	\citep{soker2011,lahav2011}.
This leads to the proposal of a `black hole fundamental plane'
   with $m_\bullet$ depending on two input quantities
	\citep{marconi2003,barway2007,hopkins2007b}.
The correlation may also take other forms,
   such as a dependence on the \cite{sersic1968} shape index
   of the stellar profile 
   \citep{graham2001,graham2007,savorgnan2013}.

More recently, it was also found 
	\citep{burkert2010,harris2011,harris2013}
   that the total number of globular clusters (GCs)
   in a galaxy correlates with the SMBH mass and with the dynamical mass
   $M_\mathrm{dyn} \approx 4 R_\mathrm{e} \sigma_\mathrm{e}^2 /G$  of the spheroidal component, 
   where $R_\mathrm{e}$ is the effective radius enclosing half of the galaxy light,
   and $\sigma_\mathrm{e}$ is the stellar velocity dispersion. 
The specific number of GCs of galaxies is not a fundamental physical property, 
   but it is a useful proxy for the total stellar mass contained in the GCs.
An interpretation of this finding 
   \citep{snyder2011}
   is that both the total mass of GCs
   and the SMBH mass correlate with the host spheroid's binding energy
   $E_\mathrm{b} \sim M_\mathrm{dyn} \sigma_\mathrm{e}^2$
	\citep[see also][]{hopkins2007a,hopkins2007b,aller2007}.   
Hence, the total number of GCs and the SMBH mass also show a correlation
   with each other.

Taking all these empirical correlations together 
   points to the presence of a general scaling relation  
   between SMBH mass, stellar mass in the host spheroid
   and the total mass/number of GCs in the galaxy.
This scaling appears straightforward to understand, at least qualitatively. 
Rapid growth of the nuclear black hole of a galaxy
   (particularly at redshifts $2 \la z \la 6$)
   might be fuelled by a massive inflow of cold gas
   towards the centre of the galaxy.
The gas inflow would trigger starbursts and the formation of new GCs. 
Numerical simulations often show
   massive gas inflows in mergers of gas-rich galaxies
   \citep[e.g.][]{hopkins2005,barnes1991,barnes1996,hernquist1989}.
This scenario expects a coeval growth of SMBH and stellar components
   (which includes the spheroid and GCs), 
   regulated by the gas supply that reaches the inner region of the galaxy,
   and ultimately primed by the merger rate 
	\citep{volonteri2009}.
The parallel growth of the SMBH and stellar component
   cannot continue indefinitely, 
   and it terminates when the gas supply ceases.   
The accretion into a black hole at super-Eddington rates 
   will emit copious radiation,
   which exerts radiative pressure on the inflowing gas, 
   leading to a massive galactic-scale outflows.
When the central black hole in a galaxy has grown to a sufficiently large mass
   (and can therefore attain a sufficiently high Eddington luminosity),
   the momentum-driven, expanding shell of the swept-up gas will achieve 
   a velocity higher than the escape velocity from the galaxy 
	\citep{silk1998,king2003,murray2005}.   
When most of the gas is expelled,
   star formation and SMBH accretion are quenched. 

However, the reality could be more complicated than described above, 
   as there is evidence that SMBH accretion
   and star formation do not always trace each other \citep{zheng2009}. 
Thus, there could be pathways (or even multiple pathways)
   of SMBH and spheroid growth 
   without invoking self-regulation \citep[e.g.][]{angles2013}  
   that lead to the SMBH scaling relations that we observe today
   \citep{zheng2013}.      
It worth noting that the duration of SMBH growth in the co-evolution scenario 
   depends on the initial mass of their seed black holes.
Some authors \citep[e.g.][]{shibata2002,volonteri2008,begelman2010}
   argued that seed black holes
   may come from direct collapse of supermassive stars, 
   which were formed directly from large-scale gas inflows
   in the DM halo. 
As such, the seed black hole mass distribution would be a function 
   of the DM halo virial temperature and the black-hole spin. 
Also, there would be an angular momentum ceiling for the DM halo,
   only below which inflows can occur and supermassive stars can form.

The existence and nature of a correlation between GCs, SMBH and dark halo 
   is not free from disputes.
GCs have a bimodal colour distribution, probably the signature of two physically distinct populations:
   younger, metal-rich red
   and older, metal-poor blue clusters
	\citep{brodie2006}.
Co-evolution of stellar populations and SMBH due to major mergers
   should produce a correlation only between red GCs (formed during
   the starburst phase and located closer to the nucleus) and SMBH
	\citep{kormendy2013}.
The correlation is indeed tighter for red GCs \citep{sadoun2012},
   but the fraction of red/blue GCs is similar for most galaxies
	\citep{burkert2010},   
   indicating some residual correlation also with the blue (old) population,
   or perhaps an initial correlation between blue GCs and seed BH. 
Intriguingly, it was recently noted
	\citep{harris2013} 
   that the relation between GC mass fraction
   (i.e., fraction of a galaxy mass that is contained in GCs)
   and galaxy mass is not a constant but has a characteristic U-shape:
   both dwarf and giant ellipticals have a larger fraction
   of baryonic mass located in GCs, than intermediate-mass galaxies.
This could be due to different rates of GC formation
   or subsequent GC destruction.
Alternatively, perhaps dwarf and giant galaxies
   have formed field stars less efficiently,
   owing to gas losses from super-winds and SMBH activity respectively.
Only in intermediate-mass systems 
   is the observed GC mass fraction a true indication
   of how much gas was initially present in the galaxy potential well.
Proponents of collisionless cold DM theories
   also invoke a scenario of gas blowouts
   to explain the differences between the simulated halo mass spectrum
   and the visible baryonic mass function,
   especially at the low-mass and high-mass ends
	\citep[e.g][]{persic1992,bell2003,read2005b,papastergis2012}.
Either way,
   it follows that the mass in GCs is determined by the amount of gas initially
   present in the (DM-dominated) potential well of a galaxy,
   and therefore there must be some correlation
   between GC mass and DM halo mass
	\citep{harris2013,georgiev2010}.
In particular,
	\cite{harris2013}
   propose a linear correlation
   with $M_\mathrm{GCs} \approx 6 \times 10^{-5} M_\mathrm{halo}$.

In summary, there are empirical hints of correlations
   between SMBHs, DM halos and GCs in galaxies
   despite the widely different scales of the three types of objects,
   but it is still not clear to what extent 
   the associations are truly intrinsic or they are mere by-products 
   of other physical processes, such as galaxy mergers.
In this work, 
   we search for physical processes that could give rise to such correlations  
   and demonstrate a physical mechanism
   that naturally links the properties of the SMBH, DM halos and GCs. 
Motivated by the extent of GC swarms
   --- rounded and far from the direct reach of
   active galactic nuclei (AGN) in normal galaxies
   ---
   we seek explanations in which
   the DM halo is the component controlling the scaling relations.

For galaxies that are
   large or small; rich or poor in baryons;
   pristine, star-forming or aged,
   observations indicate that DM haloes feature
   a kpc-scale central {\em core} of nearly uniform density,
   surrounded by outskirts where the density declines radially
   till it becomes unmeasurable at $\sim100$kpc distances
\citep[e.g.][]{flores1994,moore1994,burkert1995,salucci2000,kelson2002,
	kleyna2003,
	simon2003,gentile2004,deblok2005,thomas2005,kuzio2006,
	goerdt2006,
	gilmore2007,weijmans2008,oh2008,inoue2009,donato2009,
	deblok2010,pu2010,murphy2011,memola2011,walker2011,richtler2011,
	jardel2012,agnello2012,amorisco2012,
	schuberth2012,salucci2012,lora2012,lora2013,amorisco2013,hague2014}.
However, early theories of collisionless and non-interacting DM
   predicted steep power-law central density {\em cusps}
   and not the observed cores
	\citep[e.g.][]{gurevich1988,dubinski1991,nfw1996}.
More microphysics may be needed.
The core sizes can be set
   by the effective thermal properties of DM in equilibrium
	\citep{nunez2006,saxton2010,saxton2013},
   or temporarily by heat conduction
	\citep{kochanek2000,dave2001,ahn2005,rocha2013}.
Alternatively, one may invoke stellar and/or SMBH feedback
	\citep[see e.g.][]{navarro1996,mashchenko2006}. 
While it is conceivable to mechanically shake the halo to create a uniform core, 
   it requires certain parameter fine tuning in the feedback approach,  
   which is not always feasible in certain classes of galaxies
	\citep{gnedin2002,penarrubia2012}. 
Here, we advance a more generic theory
   that is independent of episodic astrophysical events,
   by attributing the halo structure to the innate microphysics of DM.  
Studies \citep[see e.g.][]{ullio2001,macmillan2002,
	merritt2004,zakharov2007,ghez2008,saxton2008,zakharov2010}  
   have shown the central density profile can rise locally 
   in a sharp {\em spike} in the sub-pc to pc-scale
   gravitational sphere of influence around the SMBH.
Our paper builds upon this finding,  
   allowing a direct material coupling between the halo and SMBH,  
   with a smooth transition from
   a DM density spike around the horizon of the SMBH 
   to a cored DM halo at galaxy scales,
   that in turns binds the swarm of GCs located at larger distances.
We consider constraints at both scales, 
   and show how SMBH--GC relations emerge from the SMBH--halo connection. 

We organise the paper as follows. 
Section~2 presents the model and the formulation.  
Section~3 shows the solutions,
   and in Section~4 we discuss the astrophysical implications.


\section{Model and formulation} 

\subsection{Halo model properties} 

We assume a self-gravitating,
   spherically symmetric and structurally stationary DM halo,
   in dynamical equilibrium,
   with the thermodynamics of the DM 
   described in terms of a polytropic equation of state. 
The DM in the halo is well mixed,
   without sub-halo clump structures. 
The SMBH develops from within the DM halo
   as an integral part of a self-gravitating system,  
   instead of being inserted artificially into the halo centre
   as a massive external point-like object. 
Moreover, the black hole has a physically defined horizon
   directly interfacing with the surrounding DM in the halo core.  
The gravity in the system is dominated by the DM components,
   i.e. the halo and SMBH, 
   with insignificant contribution by the baryonic components, 
   i.e. gas, stars and globular GCs.

\subsection{Equation of state of the dark matter} 
  
The equation of state of the DM takes the form
\begin{equation}
	P=\rho\sigma^2=s\rho^\gamma 
	\ , 
\label{eq.state}
\end{equation}
or equivalently
\begin{equation}
   \rho=Q\,\sigma^F
   \ , 
\label{eq.state.Q}
\end{equation}
   where $P$ is the pressure,
   $\rho$ is the density,
   and $\sigma$ is the isotropic velocity of the particles.
The quantity
   $s$ is the (pseudo-)entropy, 
   and 
   $Q\equiv s^{-F/2}$ is the phase-space density.
The adiabatic index $\gamma$ is determined by the DM microphysics. 
It is related to
   the effective thermal degrees of freedom of the dark particles $F$ via 
\begin{equation}
	\gamma = 1 +{{2}\over{F}}
	\ .
\label{eq.gamma}
\end{equation} 
Many DM scenarios entail a functionally equivalent equation of state
   (Section~\ref{s.darkmatter}).
Generally,
   $F$ describes the number of modes
   in which the microscopic energies of DM particles
   can be equipartitioned.
For translational motions in three dimensions, $F=3$.
When self-interacting DM particles
   are composite or have internal structure
   and modes of rotation, vibration and excitation
   at a comparable energy scale
	\citep{cline2013b},
   then $F>3$.
The specific heat capacity at constant volume is
   $c_\mathrm{v}\equiv Fk/2$
   (where $k$ is Boltzmann's constant)
   and the energy density is $FP/2$.
If instead DM is a sterile neutrino
   then $F\ga3$ in the degenerate halo core
	\citep[c.f. neutrino-ball SMBH:][]{viollier1993}.
If DM is a boson scalar field,
   then $F$ derives from the index of the self-coupling potential
	\citep{peebles2000}.
If DM experiences phase changes,
   then the equation of state is more complicated,
   but a polytropic law would remain a fair working approximation
   in limited ranges of temperature and density.

In principle, $Q$ and $s$ vary radially, 
   if the halo is stratified, e.g. due to a history of mergers and accretion, 
   or if dynamically significant energy exchange processes are present 
   \citep[e.g. `dark radiation,'][]{ackerman2009,fan2013a}.
In that case, buoyant stability could appear, when 
   $\mathrm{d}s/\mathrm{d}r>0$ and 
   $(F\,\mathrm{d}Q/\mathrm{d}r)<0$. 
However, we have assumed that the adiabatic DM in the halo
   does not have sub-structures.  
Thus, $Q$ and $s$ are constant for each galaxy in our calculations.

For $-2<F<10$, 
  the outer radius of the halo and total mass enclosed are finite, 
  safeguarding the existence of realistic solutions for the DM halo-SMBH system.
In this paper, we discard
   models with $F<0$,
   since they have minimum density at the centre
   and greatest densities outside
   (which seems inappropriate for galaxies).
Isolated polytropes with $F>6$ are sometimes susceptible to
   interesting dynamical instabilities
	\citep{ritter1878,emden1907,chandrasekhar1939}. 
The instability can nonetheless be moderated
   by interactions with the baryonic matter components
	\citep{saxton2013} 
   or by a confining external pressure
	\cite[e.g.][]{mcrea1957,bonnor1958,horedt1970,umemura1986}.

\subsection{Halo profile}

A realistic halo requires that the density $\rho$ falls to zero
   at a certain outer radius $R$,  
   which defines the size of the halo. 
The mass enclosed by $R$ is the total mass $M$ of the halo.  
Inside the halo, the mass distribution is the solution to
\begin{equation}
	{{{\rm d}m(r)}\over{{\rm d}r}}
	=4\pi r^2\rho(r) 
	\ ,
\label{eq.mass}
\end{equation}
   where $m(r)$ is mass contained within radius $r$. 
The gravitational field strength is given by 
 \begin{equation}
  g(r) = -{{Gm(r)}\over{r^2}}
	\ ,    
\label{eq.gfield}
\end{equation}  
  and the gravitational potential $\Phi(r)$ by   
\begin{equation}
	{{{\rm d}\Phi(r)}\over{{\rm d}r}}
	=-g(r) 
	\ . 
\label{eq.potential}
\end{equation}
The escape velocity $v(r)$ satisfies the relation
\begin{equation}
	{{{\rm d}v(r)^2}\over{{\rm d}r}}
	=2g(r)
    \ . 
\label{eq.escape}
\end{equation}

If the pressure in the halo were deficient near a central gravitating mass,
   adiabatic accretion would proceed \citep{bondi1952},
   which, conceivably, feeds the growth of the SMBH 
	\citep[e.g.][]{peirani2008b,guzman2011b,guzman2011a,pepe2012,lorac2014}.
Without losing generality we ignore the complications of accretion inflow 
   and focus on the stationary halo, 
   which is pressure-supported everywhere.
Under these conditions the velocity dispersion of DM is then given by
\begin{equation}
	{{{\rm d}\sigma(r)^2}\over{{\rm d}r}}
	={2\over{F+2}}\,g(r)
	\ .
\label{eq.hydrostatic}
\end{equation}
The DM velocity dispersion $\sigma^2(r)$ can be considered 
  as a measure of the local thermal `temperature'. 
Within the halo, this thermal temperature 
   is related to the local escape velocity and gravitational potential by 
\begin{equation}
	\sigma(r)^2 =
	{1\over{F+2}}\left[{
		v(r)^2-V^2
	}\right]
	\ , 
\label{eq.sigma}
\end{equation}  
    $V$ is the escape velocity at the outer boundary of the halo 
   ($r=R$, $\rho=0$, $\sigma=0$).  
The above expression can be obtained by carrying out an integration 
   after combining equations (\ref{eq.escape}), 
   (\ref{eq.potential}) and (\ref{eq.hydrostatic}). 
The escape velocity $V$ depends on 
  whether there is any non-DM material extending beyond the outer halo radius $R$.
Otherwise, it takes the value $V=\sqrt{2GM/R}$.

Given either the inner or outer boundary conditions,
   locating the other boundary
   is performed by numerical integration
   (Section~\ref{s.numerical}).   
We define a dimensionless gravitational compactness parameter: 
\begin{equation}
	\chi \equiv \left({
		{V}\over{c}
	}\right)^2
	={{2GM}\over{c^2R}}
	<1\ .
\label{eq.compactness}
\end{equation}
Empirical values of $\chi$ could be estimated
   from a characteristic velocity dispersion
   or mass-radius relation of
   self-bound objects.
For example,
    massive galaxy clusters have $\chi\la10^{-4}$
    ($V\la3000~\kms$);
    giant galaxies have $\chi\la10^{-6}$
    ($V\la300~\kms$);
    and faint dwarf galaxies have $\chi\la10^{-8}$
    ($V\la30~\kms$).

\subsection{Central black hole and horizon surface} 

Most galaxies are expected to possess a central black hole,  
  but observations indicate that some actually do not.
In some cases there might never occur a mass concentration
   dense enough to collapse gravitationally.
Effects such as rotational support might help avert black-hole formation
   in certain late-type galaxies
   (Section~\ref{s.ltg}).
Also, merger events could eject a SMBH from the host galaxy. 
Here however,
   we investigate only galaxies that have formed a nuclear SMBH
   and retain it in equilibrium with its DM surroundings.

The escape velocity of a test mass is $c$, the speed of light,  
   at the event horizon of a (Schwarzschild) black hole.  
If it is appropriate to consider the `formation' of a black hole 
   in this newtonian model,
   then the black hole is defined by the sphere
   where the escape velocity is $v=c$ at its surface (i.e. the horizon). 
This black hole contains a mass $m_\bullet$,
   inside a horizon radius, which is given by
   $r_\bullet\approx 2Gm_\bullet/c^2$.
In a dense DM envelope enclosing the central black hole,
   the horizon radius is larger than
   the ideal Schwarzschild value in vacuum. 
We parameterise the ratio between the horizon radius
   and the Schwarzschild radius $r_{\rm s}$ by 
   $\eta\equiv r_\bullet/r_\mathrm{s}$. 
In a fully relativistic treatment, $\eta=1$ always.
Here, the value of $\eta$ is generally of the order unity.
The mean density of the black hole is then 
\begin{equation}
	\bar\rho_\bullet
	\equiv {{3m_\bullet}\over{4\pi r_\bullet^3}}
	={{3c^6}\over{32\pi G^3 m_\bullet^2\eta^3}}
	\ .
\label{eq.dens.hole}
\end{equation}
The velocity dispersion of the DM at the horizon surface is
\begin{equation}
	\sigma_\bullet^2 = \left({
		{{1-\chi}\over{F+2}}
	}\right) \,c^2
	\ .
\label{eq.sigma.ibc}
\end{equation}
If the halo is adiabatic all the way down to the horizon surface of the central black hole,
   then from the equation of state (\ref{eq.state.Q}) we obtain 
   \begin{equation}
		{{\rho(r)}\over{\rho_\bullet}}
	=
	\left[{
		{{\sigma(r)^2}\over{\sigma_\bullet^2}}
	}\right]^{F/2}
	\ . 
\label{eq.closure}
\end{equation} 
Define a parameter $\psi\equiv \bar\rho_\bullet/\rho_\bullet$, 
    which is the density ratio of the BH
    to DM near its horizon surface.  
Then, we have 
\begin{eqnarray}
	m_\bullet
	&\hspace{-2mm}=&\hspace{-2mm}
	\sqrt{
		{{3c^6}\over{32\pi G^3}} 
	}
	\left({
		{{F+2}\over{1-\chi}}
	}\right)^{F/4}
	(\eta^3\psi\rho)^{-1/2}
	\left({
		{{\sigma}\over{c}}
	}\right)^{F/2}
\hspace{10mm}\mbox{or,}
\nonumber\\
	&\hspace{-2mm}=&\hspace{-2mm}
	\sqrt{
		{{3c^{6-F}}\over{32\pi G^3}} 
	}
	\left({
		{{F+2}\over{1-\chi}}
	}\right)^{F/4}
	{1\over{\sqrt{Q\eta^3\psi}}}
	\ .
\label{eq.m.hole}
\end{eqnarray}
Substituting any observed set of $(\rho,\sigma)$ values
   of the DM from elsewhere
   in the halo's adiabatic region
   yields an estimate of the natural mass 
   of the central compact object.
The values of $\rho$ and $\sigma$ in the above expression are local.  
They can be constrained by the observations.
The dimensionless correction factors $\psi$ and $\eta$
   are, however, obtained by numerical solution of a particular halo model.
In Section~\ref{s.core},
   we will show that,
   for the physically relevant models of polytropic DM haloes,
   the correction factors are moderate.
Appendix~\ref{s.absolute}
   expresses the mass prediction (\ref{eq.m.hole})
   in absolute physical units.

\subsection{Numerical integration scheme}
\label{s.numerical}

The radial profiles of particular polytropic haloes 
   are obtained from direct numerical integration of equations
   (\ref{eq.state}), (\ref{eq.mass}) and (\ref{eq.hydrostatic}). 
This is an initial value problem,
   with radius ($r$) as the independent variable,
   starting from either the inner boundary ($r =0$) or the outer boundary ($r=R$).
The phase-space density $Q$, (pseudo-)entropy $s$
   and thermal degrees of freedom $F$ 
   are mutually consistent constants.

We adopt the embedded eighth-order Runge--Kutta Prince--Dormand method
   with ninth-order error estimate
        \citep{prince1981,hairer2008}
	\footnote{%
        We use {\tt rk8pd} and associated routines
	from the {\sc Gnu Scientific Library}
        ({\tt http://www.gnu.org/software/gsl/}).
        }
    in our integration.
When integrating outwards from known inner values,
   we can express the differential equations in their original form.
When integrating inwards, each equation is multiplied by $-1$,
   and $-r$ is used as the independent variable.
Near the inner and outer boundaries,
   the radius is not known a priori,
   but the velocity dispersion is known exactly
   ($\sigma=\sigma_\bullet$ and $\sigma=0$ respectivey).
In those vicinities,
   it is more desirable to adopt $\sigma^2$ 
   as the independent variable in the differential equations, 
   i.e. re-expressing each quantity $y$ the equations 
   in the form of $\mathrm{d}y/\mathrm{d}\sigma^2$
   or $\mathrm{d}y/\mathrm{d}(-\sigma^2)$. 
 
In the numerical integration we first consider small steps 
   (but much larger than the round-off level) 
   until it is appropriate to switch to another independent variable
   and then continue in the same integration mode.
Doing so we can integrate accurately either
   from the outer boundary of the halo towards the SMBH horizon,
   or from the SMBH horizon to the outer boundary of the halo.


\section{Results}

\begin{figure}
\begin{center}
\includegraphics[width=74mm]{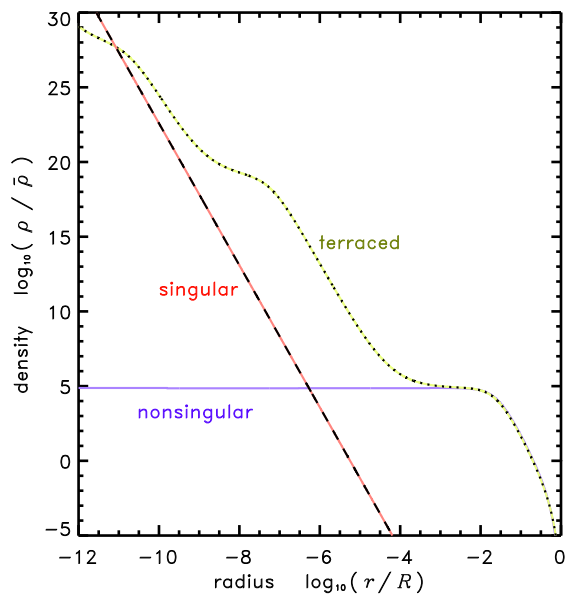}
\end{center}
\caption{%
Density profiles representing three polytrope classes.
The curves are nonsingular (solid blue),
   densely terraced (dotted green)
   and singular (dashed red) cases.
Each curve is scaled to its mean density.
They have $F=9.5$ but different $Q$ values.
}
\label{fig.classes}
\end{figure}

Before presenting the results of our DM halo-SMBH calculations, we 
   briefly review the general properties of adiabetic self-gravitating polytropic spherically symmetric bodies.   
This class of spheroids has been investigated previously, 
  but more often in the context of stars instead of larger spheroids such as galaxies or galaxy clusters. 
There are three sub-classes
   (Fig.~\ref{fig.classes})
   with these characteristics:
\begin{enumerate}
\item
\noindent 
{\bf\em Nonsingular},
    with a zero density gradient at the centre. 
    The density declines outward until reaching zero at a large radius $R$.
    The Lane-Emden spheres are examples of these \citep{lane1870,emden1907}.
    Observations show that galaxy haloes often have a uniform density core 
    with a profile resembling that of these polytropic spheres.   
\item
\noindent  
{\bf\em Singular},
    with a density spike around a massive nuclear object.
    Shallower density gradients further out resemble that of a galaxy core.
    The profile of the outer fringe is similar to
    that of the nonsingular polytropic haloes. 
\item
\noindent  
{\bf\em Terraced},
    with the radial density profile
       alternating between power-law slopes and cores,
       nested inside each other.
       The centre is, however, singular. 
   \cite{medvedev2001}
	studied terraced polytropes with $F\approx10$.
\end{enumerate}

Nonsingular polytropic spheroids can be obtained 
   by setting variables according to inner boundary conditions
   and then integrating the system of profile differential equations outwards.
In the context of DM halo-SMBH model considered in our paper
   this kind of spheroid does not provide a self-consistent description 
   for the circum-nuclear properties of the DM,
   as the central mass (SMBH) is absent.
Singular and terraced polytropic spheroids can be obtained 
   by integrating the profile ODEs inwards 
   with the outer boundary conditions   
   $M(R)=M$ and $\sigma(R)=0$. 
The central singularity gives rise to the SMBH, 
   with the horizon determined
   by setting the escape velocity equal to the speed of light. 

\subsection{Relevant solutions}

The polytropes have a nonzero compact central mass
   surrounded by a density spike, where $\rho \sim r^{-F/2}$
	\citep[e.g.][]{huntley1975,quinlan1995,ullio2001}.
The gravitational potential is keplerian near the origin,
   with $\Phi \sim r^{-1}$, 
   and the velocity dispersion peaks in the same manner, i.e.\ 
   $\sigma^2\sim r^{-1}$.
The escape velocity reaches $c$ at some sufficiently small radius.

In computing the radial profile,
   we set a fiducial outer radius, say $R=1$, 
   where $\sigma=0$,
   and choose trial values of the total mass $M$. 
This implies a specific value for the compactness parameter $\chi$.
Keeping these fixed, we test trial values of
   the phase-space density $Q$,
   and integrate the profile differential equations inwards.
If the condition (\ref{eq.sigma.ibc}) is satisfied
   then we record the conditions of that inner boundary,
   $(r_\bullet,m_\bullet,\eta,\psi)$.
If the origin is reached,
   or if a condition of $m\le0$ is encountered at any $r>0$,
   then no horizon for the central gravitating object is obtained, 
   and the trial value of $Q$ is recorded as an unphysical case.

\begin{figure}
\begin{center}
 \includegraphics[width=80mm]{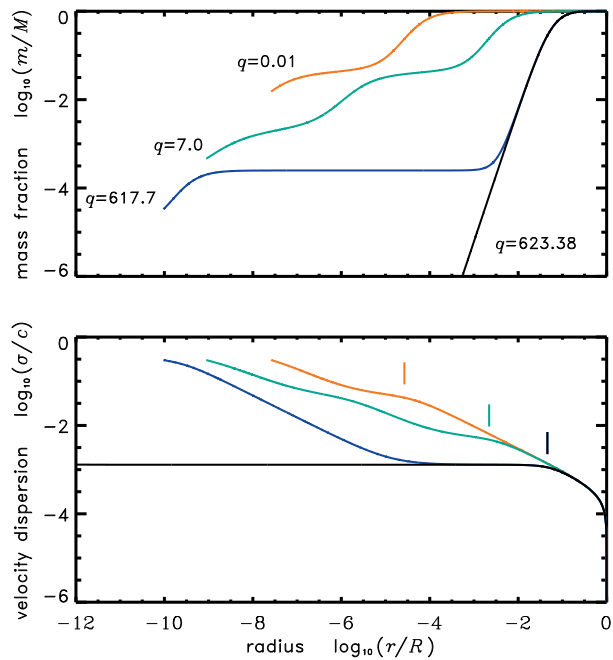}
\end{center}
\caption{%
Top:
   profile of DM + BH mass enclosed within radius $r$.
Bottom:
   the corresponding velocity dispersion $\sigma/c$ vs radius.
In all models $F=9$,
   and the compactness is galaxy-like ($\chi=10^{-6}$)
   but different phase-space densities (annotated).
The nonsingular solution is black
   ($q\equiv QV^F/\bar{\rho}\approx623.38$).
When $q=617.7$ (blue curve)
   the central mass has $m_\bullet/M=3.24\times10^{-5}$;
   the dark envelope around the horizon
   has density contrast $\psi=3.00$,
   and radius factor $\eta=2.92$.
Lower values of $q$ (higher entropy)
   give larger $m_\bullet$.
The green curve ($q=7.0$) has a terraced profile.
Vertical ticks in the lower panel mark scale radii $R_2$
   to show core sizes.
}
\label{fig.profile}
\end{figure}

Fig.~\ref{fig.profile} depicts the radial profiles of solutions
   for haloes with $F=9$
   and compactness appropriate for a galaxy
   ($\chi=10^{-6}$).
The inner tip (left) of each curve locates a horizon ($r_\bullet$);
   the outer tip (corresponding to $R=1$, which is by construction)
   is where the halo truncates itself.
Taking the outer radius to be on the order of $R\sim300$~kpc, 
   the nearly uniform core has a radius $\sim1$~kpc.
The nonsingular solution (black curve)
   is uniform at the origin.
For a model near the nonsingular limit (e.g. blue curve)
   the DM velocity dispersion (and density)
   rise at radii within a parsec.
This is the sphere of gravitational influence of the central mass.
In this particular solution,
   the dense DM envelope surrounding the horizon
   outweighs the influence of central object ($m_\bullet$)
   by almost an order of magnitude at radii $r\la10r_\bullet$.
The density contrast between the horizon and envelope is low ($\psi=3.00$).
The central mass fraction is $m_\bullet/M\approx3.24\times10^{-5}$,
   consistent with the observed ratios
   between the SMBH and their host galaxy haloes
   (e.g. under slightly different assumptions,
   the $m_\bullet$ vs $M$ results of equations (4)--(7) and Fig.5 of
	\citealt{ferrarese2002}).
For models with lower $Q$ (green and red curves)
   the inner dense-hot spike is radially larger,
   $m_\bullet$ is heavier,
   and the halo core is more compact.
At the opposite extreme (large $Q$)
   we have $m_\bullet\rightarrow0$
   and obtain the biggest possible halo core.
Its maximal mass and radius depend on $F$.
When $F$ is smaller,
   the maximal core is wide and contains much of the halo mass.
(For an incompressible fluid, $F=0$, the entire halo is a core.)
When $F$ is larger,
   the maximal core is radially smaller
   and is relatively lightweight.

\subsection{Configuration space}

\begin{figure*}
\begin{center}
\begin{tabular}{ccc}
 \includegraphics[width=160mm]{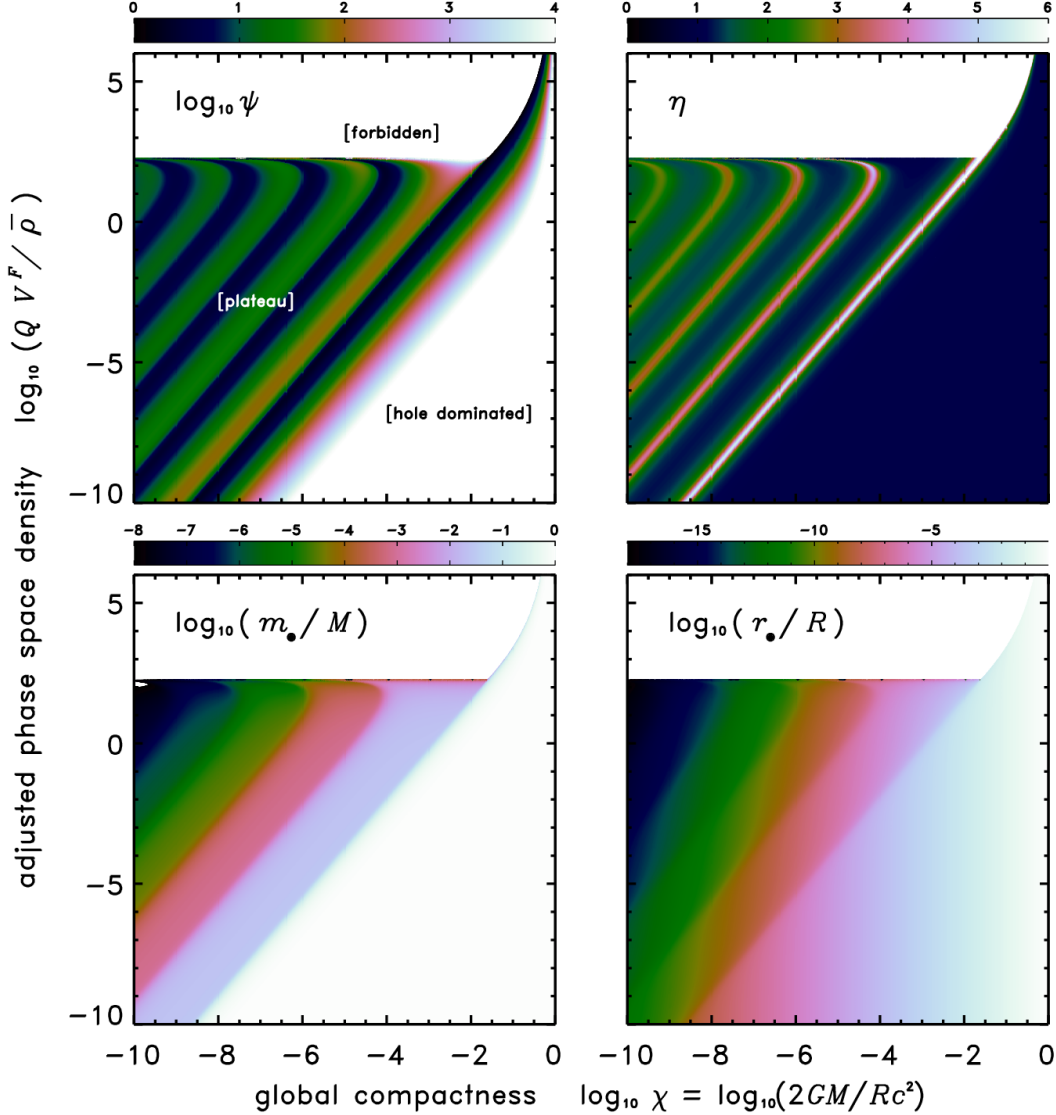}
\end{tabular}
\end{center}
\caption{%
Characteristics of the central massive object
   in a halo with $F=9.5$,
   in terms of the compactness ($\chi$)
   and adjusted phase-space density ($q=QV^F/\bar{\rho}$).
The upper panels show
   the density contrast at the horizon
    ($\psi=\bar{\rho}_\bullet/\rho_\bullet$);
   and the correction to Schwarzschild radius
   ($\eta=r_\bullet/r_\mathrm{s}$).
Lower panels show
   the mass and radius fractions
   ($m_\bullet/M$ and $r_\bullet/R$).
}
\label{fig.qplane}
\end{figure*}

Particular radial profiles
   can be obtained for choices of $(\chi,Q)$
   across a two-dimensional configuration space at fixed $F$.
This task can be
   wrapped within a root-finding routine
   or an amoeba-like minimiser,
   seeking a specific or optimal value of
   any desired property of the central object
   (e.g. $m_\bullet/M$ or $\psi$).
We explore the $(\chi,Q)$ plane numerically at high resolution.
Fig.~\ref{fig.qplane}
   maps the varying properties of the central object,
   for $F=9.5$ halo models.
For clarity of presentation,
   the vertical axis value is
   a dimensionless adjusted version of the phase-space density,
\begin{equation}
	q\equiv Q V^F / \bar{\rho}
\end{equation}
   where $\bar{\rho}=3M/4\pi R^3$
   is the mean density of the system,
   and $V$ is the surface escape velocity.
The four panels show results for:
   the horizon density contrast ($\psi$);
   the horizon radius correction ($\eta$);
   the central object's fractional mass ($m_\bullet/M$);
   and its radius ($r_\bullet/R$),
Several distinct domains appear.
The top-left panel labels these domains:
\begin{enumerate}
\item
\noindent{\bf\em forbidden zone:}
For sufficiently high $q$ there are no self-consistent solutions.
The halo is too dense and cold to reach the assumed outer radius $R$.
\item
\noindent{\bf\em border zone:}
For $q$ slightly below the forbidden zone,
   there is a thin domain of solutions
   with extremely high or low $\psi$ values
   (and steep gradients of $\partial\psi/\partial q$).
The upper edge of the border
   is where the nonsingular solutions occur
   ($m_\bullet/M\rightarrow0$ and $r_\bullet/R\rightarrow0$,
   which cannot describe a galaxy hosting a SMBH).
\item
\noindent{\bf\em moderate plateau:} 
If $6<F<10$ then there is a domain of $q$ values below the border,
   where $m_\bullet/M$ and $r_\bullet/R$ are small but finite
   (and astronomically significant).
The envelope density is non-negligible compared to
   the mean density of the black hole ($\psi\la100$).
This plateau zone is more extensive in $q$ (or in $Q$)
   if $\chi$ is small
   (systems with low escape velocities).
Viewed in $(\chi,q)$ or $(\chi,Q)$ planes,
   the plateau is roughly triangular.
Terraced haloes occur here.
\item
\noindent{\bf\em valleys:}
Within the plateau,
   there are local minima in $\psi$,
   coinciding with spikes in $\eta$.
This implies gradual density continuity between
   the BH and its immediate dark envelope.
Valleys are more numerous for smaller $\chi$.
\item
\noindent{\bf\em hole-dominated:} 
At $q$ values lower than the plateau zone,
   the black hole mass becomes dominant,
   $m_\bullet/M\rightarrow1$.
The halo is relatively tenuous:
   $r_\bullet/R$ remains small.
The black hole is effectively decoupled
   from the density and pressure of its diffuse surroundings.
The density contrast $\psi$ rises by orders of magnitude,
   and the gradient $\partial\psi/\partial q$ is steep.
\end{enumerate}

\begin{figure*}
\begin{center}
\begin{tabular}{ccc}
 \includegraphics[width=168mm]{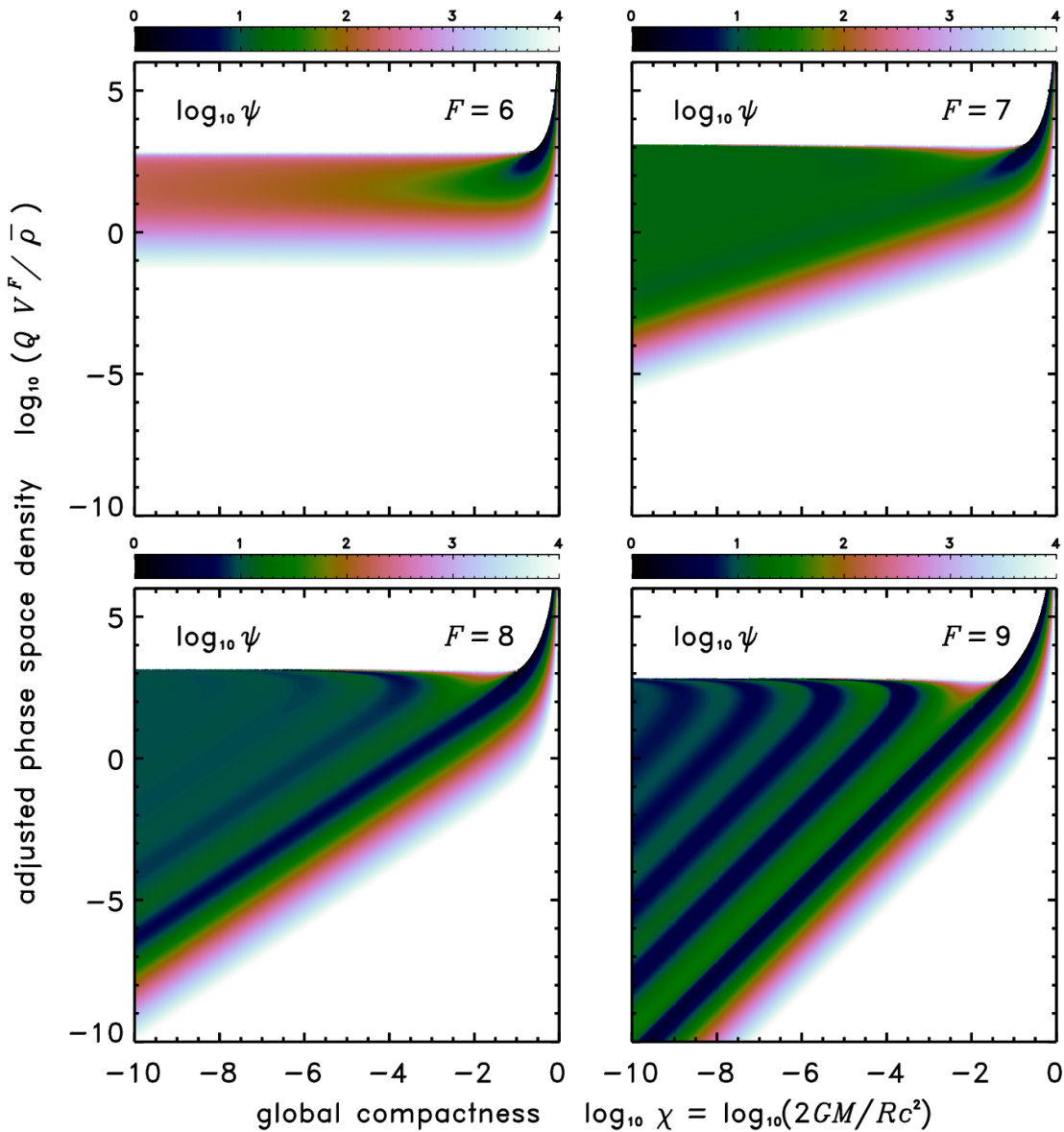}
\end{tabular}
\end{center}
\caption{%
Maps of the contrast between DM envelope and SMBH
   ($\log_{10}\psi$ shaded)
   across the $(\chi,q)$ configuration space.
The $\psi$-valleys are the dark streaks.
Panels depict haloes with $F=6, 7, 8, 9$ as annotated.
}
\label{fig.valleys}
\end{figure*}

The plateau and $\psi$-valleys
   only exist for haloes with $6<F<10$.
For $0\le F\le6$, the transition from nonsingular border
   to the hole-dominated domain is much narrower than 1\,dex in $q$.
For $6<F<10$,
   the plateau becomes wider with increasing $F$,
   until the plateau vanishes suddenly around $F=10$
   (infinite profiles including the classic \citealt{plummer1911} model).
Fig.~\ref{fig.valleys}
   depicts the $\psi$ landscape of the plateau
   in the $(\chi,q)$ plane,
   for various equations of state ($F=6, 7, 8$ and $9$).
The $\psi$-valleys are conspicuous diagonal stripes.
The valleys are more numerous for greater $F$.
For large-$F$ models, the $\psi$-valleys coincide
   with steps in the ratio $m_\bullet/M$
   (Fig.~\ref{fig.qplane}, compare left panels).
For lower $F$, the steps are less distinct
   (gradients $\partial(m_\bullet/M)/\partial q$
   and $\partial(m_\bullet/M)/\partial \chi$ are steadier).
Across most of the plateau, the $\psi$ contours (such as the valleys)
   are approximately parallel to contours of $m_\bullet/M$.
As noted below (Section~\ref{s.energy})
   the $\psi$-valleys are locally energetically favoured states.
Haloes in $\psi$-valleys have varied properties:
\begin{enumerate}
\item
\noindent 
In the valley at lowest $Q$,
   the BH is obese ($m_\bullet/M\ga0.1$)
   with only a tenuous halo.
This is unrealistic for a galaxy.
\item 
\noindent 
Valleys at intermediate $Q$: dark diagonals in Fig.~\ref{fig.valleys}.
Running along these valleys
   keeps $m_\bullet/M$ nearly constant,
   resembling a \cite{magorrian1998} relation.
\item 
\noindent 
Valleys near the nonsingular border
   have branches and irregular $\psi$-topography.
Values of $m_\bullet/M$ are lowest here.
\end{enumerate} 

For larger $F$,
   the valleys reach lower values of $m_\bullet/M$ at any given $\chi$.
For $\chi\approx10^{-6}$
   the lowest valley haloes with $F=8$ and $F=9$
   give $m_\bullet/M\sim10^{-3.5}$ and $\sim10^{-4.5}$ respectively.
This range may be consistent with observed SMBH--bulge relations
   if the DM halo is $\sim10^1$ times the baryonic mass.
Thus, on the one hand,
   when $7\la F<10$ some of the $\psi$-valleys
   are consistent with realistic SMBH masses relative to the host galaxy.
Conversely, assuming these theories of $F$,
   we predict that some galaxies host SMBH
   in low-$\psi$ configurations:
   the dark envelope is dense near the horizon.
At least in the present newtonian model,
   the edge of such a SMBH is blurry.
This deserves further investigation through general relativistic calculations.

\subsection{Relation of central mass to halo core}
\label{s.core}

Formally the configuration space at fixed $F$ is two-dimensional,
   with gravitational compactness and phase-space density parameters
   $(\chi,Q)$ or $(\chi,q)$.
In practical applications,
   the compactness parameter
	($\chi=2GM/Rc^2$)
   may be difficult to estimate,
   since it depends on the total mass $M$
   contained within the dark halo truncation radius $R$,
   which is not directly observable.
It may be more useful to specify models
   in terms of quantities pertaining to
   the measurable DM core.
If the local index of the density profile is
\begin{equation}
	\alpha\equiv -{{\mathrm{d}\ln\rho}\over{\mathrm{d}\ln r}}
	\ ,
\end{equation}
   then we can annotate slope-radii
   ($R_\alpha<R$) at a standard chosen $\alpha$.
The radii $R_\alpha$ can be multi-valued
   (in terraced haloes)
   and this is more likely when $F$ is larger.
We shall define the DM core to be the region enclosed by
   the outermost locations where $\alpha=1,2,3$,
   i.e.  $R_1 < R_2 < R_3 < R$.
The mass contained within these radii satisfies
   $m_\bullet\ll M_1 < M_2 < M_3 < M$.
The gravitational compactness of the core can be written as
   $\chi_\alpha\equiv -2\Phi_\alpha/c^2$.
Similarly, we might define the core at the half-mass radius $R_m$
   and potential $\Phi_m$
   (where $m(R_m)={\frac12}M$),
   with core compactness $\chi_m=-2\Phi_m/c^2$.
In our discussions below,
   we can abbreviate the core compactness $\chi_\mathrm{c}$ 
   standing for $\chi_1$, $\chi_2$, $\chi_3$ or $\chi_m$.
The particular choice does not change the qualitative conclusions.
These notations disregard the tenuous outskirts of the halo,
   which in any case are difficult to measure astronomically.

\begin{figure*}
\begin{center}
\begin{tabular}{ccc}
 \includegraphics[width=160mm]{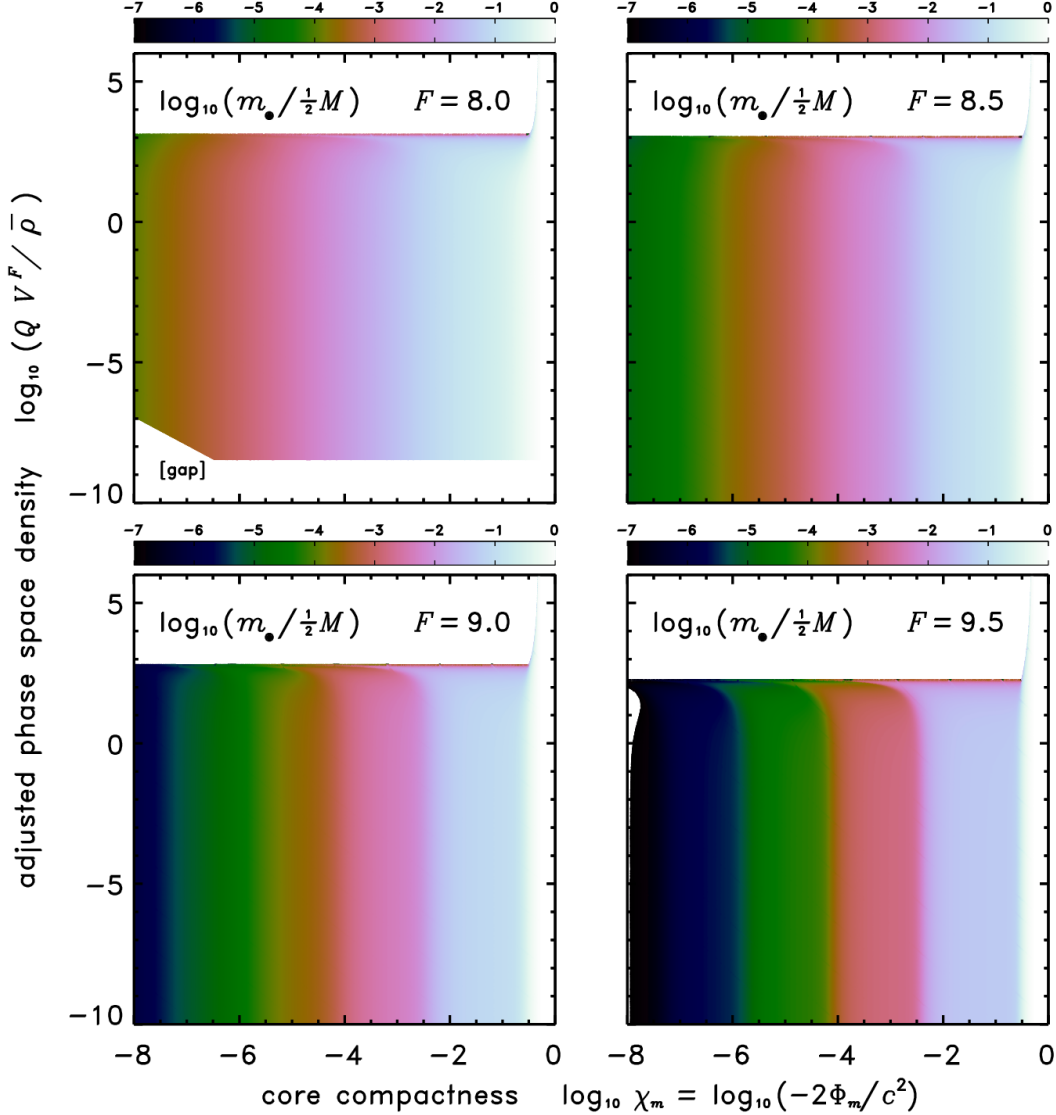}
\end{tabular}
\end{center}
\caption{%
Mass fraction of the central object
   (shaded)
   in relation to the DM core,
   for various values of $F$ (as annotated).
Use of core compactness
   (e.g. $\chi_\mathrm{c}=\chi_2$, $\chi_3$ or $\chi_m$)
   instead of global compactness $\chi$
   reveals a projection in which the model properties
   are insensitive to $q$ except near the upper border (nonsingular solutions).
}
\label{fig.phiplane}
\end{figure*}

When the $(\chi,q)$ plane transforms to
   $(\chi_\mathrm{c},q)$,
   the plateau region straightens from a wedge
   with diagonal stripes
   to a rectangular region with vertical stripes
   (see Fig.~\ref{fig.phiplane}).
Except near the upper $q$ border (the nonsingular limit)
   the models of fixed $\chi_\mathrm{c}$
   are almost independent of $q$.
The two-dimensional parameter-space is almost (but not quite)
   reduced to a one-dimensional space
   in terms of the core compactness.

\begin{figure*}
\begin{center}
\begin{tabular}{ccc}
 \includegraphics[width=135mm]{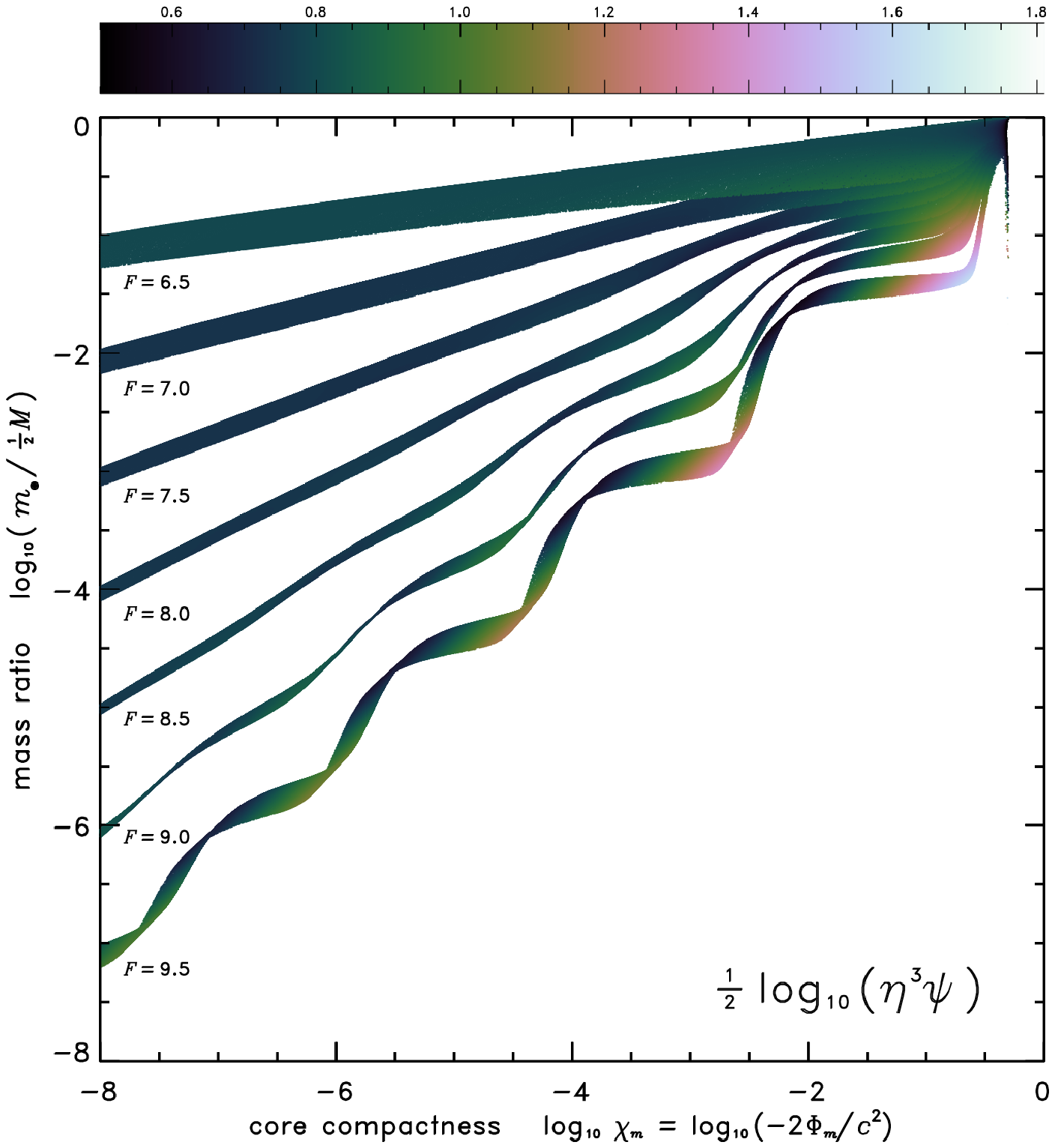}
\end{tabular}
\end{center}
\caption{%
Ratio of central mass ($m_\bullet$)
   to the mass within the halo core
   (shown as the half-mass, $M_\mathrm{c}={\frac12}M$)
   as a function of the core compactness.
From top to bottom,
   the ribbons represent the `plateau' domain (terraced halos)
   in cases of $F$ as annotated.
Colours indicate the correction term in equation (\ref{eq.m.value}).
To exclude the nonsingular border
   (where $q=q_\mathrm{ns}(F,\chi)$)
   we only plot data with
   $q<q_\mathrm{ns}/6^{F/10}$.
}
\label{fig.ribbons}
\end{figure*}

Fig.~\ref{fig.ribbons}
   shows the variation of $m_\bullet/M_\mathrm{c}$
   with respect to core compactness,
   for equations of state with $F=6.5, 7.0, 7.5, 8.0, 8.5, 9.0$.
Each ribbon depicts the entire plateau region,
   apart from the border strip.
The physically uninteresting `hole-dominated' region
   hides in the top-right point where $m_\bullet\approx{M}$.
The thinness of the ribbons in this projection
   shows how $q$ becomes inconsequential compared to $\chi_\mathrm{c}$.
The nonsingular solutions (not shown)
   have smaller $m_\bullet/M_\mathrm{c}$ for given $\chi_\mathrm{c}$:
   down to arbitarily small values as $q$ approaches its maximum.
In Fig.~\ref{fig.ribbons}
   they occupy the region of the $(\chi_\mathrm{c},m_\bullet/M_\mathrm{c})$ plot
   below the ribbon of given $F$.
Thus each ribbon represents the maximum possible $m_\bullet$
   hosted within a DM halo core of given compactness.

Now we can interpret a characteristic velocity dispersion
   of tracer objects in the core region,
   $\sigma_\mathrm{gc}\propto c\sqrt{\chi_\mathrm{c}}$.
Observable velocity dispersions of the globular cluster swarm
   should match this value to within a factor of a few.
For an assumed halo $F$ and known $\sigma_\mathrm{gc}$,
   one can estimate $\chi_\mathrm{c}$
   and infer a narrow range of possible $m_\bullet/M$
   (if the galaxy is in the `plateau' regime)
   or an upper limit on $m_\bullet/M$
   (if it is a nearly nonsingular case).
An estimate $\sigma\propto\sigma_\mathrm{gc}$
   can be substituted in equations
   (\ref{eq.m.hole}) and (\ref{eq.m.value}).
Estimates should be most robust
   for galaxies where GCS projected $\sigma_\mathrm{gc}$
   appears nearly constant within the DM core
	\citep[e.g.][]{cote2003,bridges2006,norris2012,napolitano2014}.

The shading of the ribbons in Fig.~\ref{fig.ribbons}
   shows the horizon correction factor
   $\log_{10}\sqrt{\eta^3\psi}$
   that applies in equation (\ref{eq.m.hole})
   when predicting SMBH masses in physical units.
For the $F>6$ plateau halo models,
   the variation of this term is small.
For galaxy compactness ($\chi_\mathrm{c}\ll10^{-4}$)
   the correction factors vary by less than $0.6$\,dex.
The variation is smaller for low $F$.
Thus equation (\ref{eq.m.hole})
   is robust enough to apply approximately,
   even when $Q$ is unobservable.


\section{Discussion}

\subsection{Preferred solutions and $m_\bullet$}

Although the $\chi_\mathrm{c}$ core representation
   simplifies the projected configuration space,
   the present spherical halo model still has two free parameters:
   the compactness $\chi$,
   and some measure of the orderliness (such as $Q$ or $q$).
Given the apparent simplicity of the empirical relations
   between $m_\bullet$ and host galaxy properties,
   it is worth seeking a simple causal explanation.
Is there any physical principle
   that constrains $q$ as a function of $\chi_\mathrm{c}$ or $\chi$?
A satisfactory model would involve a simple intuitive rule
   involving instantaneous halo properties,
   without complexities 
   involving fine-tuning processes, such as those invoked in feedback scenarios, 
   local contingencies or accidents of evolutionary history.   
Here we shall discuss some conceivable rule-of-thumb explanations.

\subsubsection{cosmic density}
\label{s.virial}

In the theory and simulations
   of cosmological collapse of collisionless haloes,
   galaxy-like objects lack a clearly defined outer boundary surface. 
Instead they are described in terms of a virial radius
   containing some multiple of the cosmic critical density
   (e.g. $\rho_\mathrm{v}=100\rho_\mathrm{crit}$
	with $\rho_\mathrm{crit}
	\approx9.2\times10^{-30}~\mathrm{g}~\mathrm{cm}^{-3}$,
	when $H_0\approx70\kms\,\mathrm{Mpc}^{-1}$;
	e.g. \citealt{hinshaw2013}).
It is not obvious whether haloes in polytropic theories
   would also separate from the cosmic background
   at a standard density. 
If they do,
   then their individual radii and compactness are linked,
\begin{equation}
	R=\sqrt{
		{3c^2\chi}\over{8\pi G\rho_\mathrm{v}}
	}
\end{equation}
   and the corresponding mass is
   $M=\chi c^2 R/2G \propto\sqrt{\chi^3/\rho_\mathrm{v}}$.
Thus, the dimensionless parameter $\chi$
   is linked to empirical properties of each galaxy.
Halo star and GC velocity dispersions would scale as $\sigma\sim\sqrt\chi$.
Local densities would scale in proportion to the standard $\rho_\mathrm{v}$.
The internal mass distribution still depends non-trivially upon $q$ or $Q$
   however,
   which determines whether 
   a particular galaxy halo is nonsingular,
   highly singular,
   or any condition in between.
If $\eta=\eta(\chi,q)$ were weakly dependent on $q$
   then equation~(\ref{eq.m.hole})
   would imply a power-law correlation,
   $m_\bullet\sim \sigma^{F/2}$.
If $\eta=\eta(\chi,q)$ also has non-trivial variations,
   the assumption of a universal virial density $\rho_\mathrm{v}$
   would not predict $m_\bullet$ tighter than the ribbon relations in
   Fig.~\ref{fig.ribbons}
At least one extra principle is needed.

\subsubsection{halo entropy}

The total entropy of the DM in the halo is
\begin{equation}
	S=-Nk\ln(Q/Q_0)
\end{equation}
   where the constant $Q_0$ depends on universal particle properties,
   $N=(M-m_\bullet)/\mu$ is the number of DM particles,
   and $\mu$ is the particle mass.
The event horizon also contributes entropy,
   $S_\bullet\approx\pi k(r_\bullet/l_\mathrm{P})^2$
   where $l_\mathrm{P}$ is the Planck length
   \citep{bekenstein1973}.
When models are normalised
   (Appendix~\ref{s.scaling})
   to the same total mass $M$,
   the total entropy is
\begin{equation}
	S =
	4\pi k\left({
		{{M}\over{m_\mathrm{P}}}
	}\right)^2
	\left({
		{{m_\bullet}\over{M}}
	}\right)^2
	\eta^2
	-{{M}\over{\mu}}\left({
		1-{{m_\bullet}\over{M}}
	}\right)
	k\ln\left({{Q}\over{Q_0}}\right)
	\ .
\label{eq.entropy}
\end{equation}
   where $m_\mathrm{P}$ is the Planck mass.
The left (horizon) term of (\ref{eq.entropy}) dominates if
    $\mu\gg m_\mathrm{P}^2/M$
and vanishes if
    $\mu\ll m_\mathrm{P}^2/M$.
At fixed $\chi$,
   the right term is monotonic in $Q$,
   and so is the left term, except subtle wrinkles
   within 1~dex of the nonsingular border.
Maximal entropy prefers
   a maximally massive BH
   with only a tenuous dark envelope.
Realistic SMBH scaling relations
   cannot derive from a simple entropic principle.

\subsubsection{energetic constraints}
\label{s.energy}

For the $F>6$ scenarios,
   some mass profiles are energetically more or less favourable,
   depending non-trivially on the system parameters.
For fixed $M$ and $\chi$,
   the gravitational potential energy $|W|$
   and total energy are extremal
   at the $\psi$-valley where $q$ is lowest.
This is the valley where $m_\bullet/M\ga0.1$, which is excessive.
The black hole mass is significant compared to the envelope,
   but not dominant.
This solution is energetically favoured because
   DM is mostly concentrated deep in the potential well.
At lower $q$, the black hole dominated profiles ($m_\bullet/M\approx1$)
   are less energetically favourable
   because there isn't much matter in the tenuous halo.
In the medium-$q$ plateau domain,
   the configurations are less energetically favourable
   because the mass is less concentrated.

However,
   the other $\psi$-valleys
   (where the $m_\bullet/M$ ratios are more astronomically realistic)
   are subtle local extrema of $|W|$.
In energetic terms, these states may be {\em locally} prefered
   to adjacent configurations in $(\chi,q)$ space.

It isn't obvious whether or not these energetically favourable states
   are effective attractors in galaxy halo evolution.
An evaluation of realistic evolutionary tracks in $(\chi,q)$-space
   might require Monte Carlo simulations
   that apply hierarchical mergers
   to an initial population of primordial mini-haloes.
As in toy-model studies of SMBH demographics
	\citep[e.g.][]{yu2002}
   it would be necessary to assume whether
   adiabatic agglomeration
   or mass-energy conservation
   takes priority during mergers, flybys and fission events.
Dark shocks and mixing would introduce inelastic and dissipative factors. 
These issues are non-trivial and deserve a separate investigation.

\subsubsection{landscape of $\psi$}

The density ratio of the central object to its envelope, $\psi$,
   is a diagnostic of the halo solutions.
One might wonder whether a sensible physical condition involving $\psi$
   might select the astronomically realistic models.
Large values of $\psi$ imply a central object
   with a high density contrast to its surroundings.
The rare cases with $\psi<1$ are perhaps unnatural
   as they would imply an overdense inner envelope,
   and a density inversion in whatever primal object
   formed the SMBH seed in the first place.
Small values of $\psi\ga1$ are of special interest,
   as they imply systems where the inner DM envelope
   is comparable to the mean density of the central object.
In some sense, this implies a SMBH
   that is maximally blended and coupled with the galaxy halo.

An inner condition
   $\partial m_\bullet/\partial r_\bullet = {\mathrm d}m/{\mathrm d}r$
   would describe a seamless continuity
   between the SMBH and its envelope.
If the horizon occurred at the Schwarzschild radius
   ($r_\bullet\approx r_\mathrm{s}$ and $\eta\approx1$)
   the optimal continuity condition would imply $\psi\approx3$.
When the envelope is massive enough that $\eta\neq1$
   by a significant amount,
   the seamless condition will prefer another value of $\psi$
   (depending on the actual non-Schwarzschild
   $\partial m_\bullet/\partial r_\bullet$ at fixed $\chi$).
Ideally that rule should be calculated from the numerical maps of
   $\partial m_\bullet/\partial q$ and $\partial r_\bullet/\partial q$.
We would expect the special value of $\psi$
   to be a number of order three or unity:
   probably near the minima in the $\psi$-valleys,
   and not in the large-$\psi$ regions outside the plateau.

The seamless envelope condition is theoretically interesting,
   but what would it imply about SMBH formation and growth?
It might be the natural outcome if DM accretion was
   the main mass supply to the black hole,
   either through gradual, adiabatic contraction
   or violent, supernova-like implosions.
If the subtle energetic preference for $\psi$-valleys
   were an evolutionary attractor,
   this qualitative picture
   might become something more quantitatively predictive.

\subsubsection{summary}

If haloes share a cosmologically determined mean density,
   then their individual masses are functions of $\chi$,
   but the variation of internal structure
   means this ansatz does not provide unique predictions for $m_\bullet$.
Entropy maximisation ideally favours high $m_\bullet/M$,
   which could not describe a realistic galaxy.
Gravitational energy also favours models with $m_\bullet$ too large,
   but there is a subtler preference for moderate-$m_\bullet$ profiles
   in the $\psi$-valleys of the $(\chi,q)$ space.

If galaxies tend to evolve to minimise $\psi$
   then this implies that a relativistic dark envelope
   surrounds the SMBH horizon,
   where local densities of DM
   could be significant compared to the SMBH mean density.
In that case the preferred configurations include:
   tracks where $m_\bullet/M$ is almost independent of $\chi$
   (resembling Magorrian relations);
   a track of minimum $q$ with unrealistic $m_\bullet/M$;
   and low $m_\bullet/M$ cases near the maximum $q$
   (nonsingular border, minimal entropy).
Detailed scaling relations depend on $F$ sensitively.

When the halo is described in terms of properties of its DM core,
   the plateau region of the parameter space simplifies considerably.
In these terms, the $m_\bullet/M_\mathrm{c}$ ratio
   falls within a narrow ribbon that depends on $F$ and $\chi_\mathrm{c}$
   but only weakly on $q$.
The valley and inter-valley solutions
   shrink together into the same projected region.
The ribbons are thinner than the present
   scatter in observational $m_\bullet$ values,
   so it is {\em practically} almost a one-dimensional
   $m_\bullet$--$\sigma$ scaling relation.
When nearly nonsingular models are allowed,
   the ribbons in Fig.~\ref{fig.ribbons}
   become strict upper limits on $m_\bullet/M_\mathrm{c}$.
If we can link the velocity dispersion of halo tracers
   --- such as globular clusters ---
   to the core compactness
   ($\chi_\mathrm{c}=\chi_1$, $\chi_2$, $\chi_3$ or $\chi_m$)
   then realistic $m_\bullet$ vs $\sigma$ relations emerge.
These relations may have some intrinsic scatter,
   due to the model dependencies of $\sqrt{\eta^3\psi}$
   envelope correction factor.
This factor reintroduces some $Q$-dependence,
   though it is subtle:
   for $6.5\le F\le9.5$
   and $\chi_\mathrm{c}>10^{-8}$
   we find
   $0.5<{\frac12}\log_{10}(\eta^3\psi)<1.81$
   across the plateau region
   (terraced haloes, as coloured in Fig.~\ref{fig.ribbons}).
For galaxy-like compactness
   $\chi_\mathrm{c}\ll10^{4}$,
   $0.5\la{\frac12}\log_{10}(\eta^3\psi)<1.1$

\subsection{Comparison to observed SMBH}

Empirically,
   the heaviest known ultramassive black holes
   amount to a few times $10^{10}m_\odot$
	\citep{mcconnell2011,mcconnell2012,vandenbosch2012}.
The smallest confirmed SMBH are a few times $10^5m_\odot$,
   residing in bulgeless discs and dwarf galaxies
	\citep{filippenko2003,barth2004,peterson2005,shields2008,
		seth2010,reines2013}.
For the central black hole of a massive elliptical galaxy,
   $m_\bullet=10^9m_\odot$
   and the Schwarzchild radius 
   $r_\mathrm{s}\approx2.95\times10^{14}~\mathrm{cm}
	\approx10^{-7}~\mathrm{kpc}$.
For a realistic galaxy-sized halo,
   this implies
   $r_\bullet/R\la10^{-9}$;
   and $m_\bullet/M\la10^{-3}$ or $\la10^{-4}$
   assuming a large galaxy with $\chi\sim10^{-6}$.
We adopt these values as a benchmark.

For haloes with
   $0\le F\le6$,
   astronomically realistic ratios of $m_\bullet/M$
   only exist very close to the border of nonsingular models.
A small relative amount of heating
   (e.g. dissipative effects of a tidal flyby)
   could induce a significant jump in $m_\bullet$,
   unless the compactness $\chi$ also changes.
In the thin band of $(\chi,q)$ states where $F\le6$
   is compatible with SMBH scaling trends,
   the density contrast between the hole and its dark envelope
   tends to be immense ($\psi\gg10^{10}$).
This means that when $F\le6$,
   an astronomically realistic BH is much denser than its surroundings,
   and effectively decoupled from the ambient halo pressure.
It would be necessary to invoke elaborate, mundane non-DM physics
   to explain the observed correlations.
The $6<F<10$ regime however enables
   observationally plausible $m_\bullet/M$ values
   throughout the small-$\psi$ plateau and valleys,
   as well as near the nonsingular border.
Many orders of magnitude are available in $q$ and pseudo-entropy $s$.
Incrementally heating a galaxy halo
   need not be catastrophic for SMBH growth.

Observational constraints are met in the $\psi$-plateau
   when $6<F<10$ (for $r_\bullet/R$)
   and $7.5\la F<10$ (for $m_\bullet/M$,
   according to our Fig.~\ref{fig.ribbons}).
This then is an observationally favoured range
   of DM microphysics.
We typically find $\psi<100$ in the best models.
This entails a dark envelope with gravitationally significant density
   near the event horizon.
This envelope declines radially as a nuclear `spike',
   though more steeply than the low-$F$ spikes of previous modelling
	\citep{gondolo1999,mouawad2005,hall2006,zakharov2010}.
For large $F$, 
   the spike's steepness
   makes the combined DM envelope plus BH
   appear (from afar) as if it were a more massive black hole.

To an accuracy comparable to the present observational scatter,
   it is useful to represent the mass trend as a power-law,
   $m_\bullet/M_\mathrm{c}\sim M_\mathrm{c}^{\beta-1}$.
If we assume that galaxy haloes share a nearly universal cosmic mean density
   (Section~\ref{s.virial},
   $M_\mathrm{c}\propto\sqrt{\chi_\mathrm{c}^3/\rho_\mathrm{v}}$)
   then the expectation is
   $m_\bullet/M_\mathrm{c}\sim\chi_\mathrm{c}^{3(\beta-1)/2}$.
In our numerical results, the domain $6<F<10$ ensures $1<\beta<2$;
   while $F<6$ only gives solutions near the nonsingular limit ($\beta=1$).
Observations of $m_\bullet$
   in local galaxies and AGN \citep{laor2001}
   show $\beta=1.54\pm0.15$.
In our
   Fig.~\ref{fig.ribbons}
   this would correspond to slopes of
   $\partial\ln(m_\bullet/M_\mathrm{c})/\partial\ln\chi_\mathrm{c}=0.81\pm0.23$,
   which graphically is consistent with the ribbons of higher $F$ cases.
\cite{bandara2009}
   modelled the strong gravitational lensing effects
   of a set of elliptical galaxies
   that also have $m_\bullet$ estimates.
They found a correlation that
   implies $\beta=1.55\pm0.31$ or $\beta=1.57\pm0.39$
   depending on their fitting methods.
Our equivalent ribbon slopes would be
   $\partial\ln(m_\bullet/M_\mathrm{c})/\partial\ln\chi_\mathrm{c}
	=0.82\pm0.47$ or $0.86\pm0.59$.
These constraints are lax,
   but would seem to prefer $F\ga7$.

In our model, the predicted ratios $m_\bullet/M$
   refer to $M$ as the halo mass,
   not the stellar bulge ($M_\bigstar$).
Peculiar galaxies observed with high $m_\bullet/M_\bigstar$
   \citep{bogdan2012,vandenbosch2012}
   could be normal products of the SMBH-halo relationship,
   but impoverished in stars and gas for some other reason.
Alternatively, if they are genuinely overweight in $m_\bullet/M$ terms,
   they might be high-entropy outliers:
   low $q$ or high $\chi_\mathrm{c}$
   due to an unlucky history of tidal buffeting or other halo heating processes.

Our model
   also has implications for the presence of
   intermediate mass black holes
   (IMBH; $10^3\la m_\bullet/m_\odot\la10^6$)
   in the least massive systems.
Based on velocity dispersions, escape velocities and tidal radii,
   ultra-compact dwarf
   and faint dwarf galaxies
   could have compactness parameters $\chi\la10^{-7}$.
If they bind substantial amounts of DM
   then the $\psi$-plateaus of $F>6$ haloes
   set upper limits on $m_\bullet/M_\mathrm{c}$ that are rather low
   (left extreme of Fig.~\ref{fig.ribbons}).
In a system amassing $M=10^6m_\odot$,
   the `plateau' configurations of $F=7,8,9$ haloes
   with $\chi\approx10^{-8}$
   predict a maximum central object of
   $m_\bullet\approx10^{4}, 10^{2}$ and $10^{0}m_\odot$ respectively.
For objects with $\chi\approx10^{-7}$
   these $F=7,8,9$ models give
   $m_\bullet\approx10^{4.5}, 10^{2.5}$ and $10^{0.5}m_\odot$ respectively.
This object could be a stellar black hole, rather than an IMBH.
If there is a non-trivial central stellar density,
   then the predicted central mass is lost amidst stellar granularity,
   and the model breaks down.
Even if an IMBH were formed,
   there are plausible processes that might remove it:
   the `gravitational rocket' effect during high-spin black hole mergers;
   random walks due to scattering in dense stellar environments;
   random walks due to momentary imbalances between the thrusts of two jets
   during a gas accretion episode.
The rarity or non-observation of IMBH in dwarf galaxies and GC
   is unsurprising.

\subsection{Possible observational tests of the dark matter envelope}
\label{s.envelope}

The presence of a DM envelope around a black hole
   contributes to the gravitational potential, 
   which will produce observational consequences.  
Here we list a few examples.

(i) The gravitational potential of the DM envelope 
   will cause stellar orbits to deviate from the Keplerian orbits
   that are expected for motion around
   a bare spherically symmetric gravitating object
	\citep{rubilar2001,hall2006,mouawad2005,zakharov2007,ghez2008,will2008,
	zakharov2010,iorio2011}.
A possible means to detect this deviation is timing observations of pulsars, 
   if present, around the central black holes in nearby galaxies
	\citep{wex1999,pfahl2004,kramer2004,liu2012,singh2014}.

(ii) Stars with non-circular orbits
   traversing the DM envelope around a black hole 
   would experience a gentler tidal-force gradient
   than around a bare BH of equal total mass.
In a stellar tidal disruption process
	\citep[see][]{rees1988,komossa2002,bloom2011,saxton2012}
   the stellar debris tracks would have morphologies  
   different to those resulting from a rapid change in the tidal force field. 

(iii) AGN are powered by accretion of gas into a massive black hole. 
The inner accretion disc region
   unleashes most of the accretion power,
   in the form of radiation and outflows.
For objects orbiting around a black hole, 
   there is a well-defined innermost stable circular orbit (ISCO).
This orbit is assumed to be the inner boundary of the accretion disc,
   because beyond that, the inflow matter plunges towards the horizon
   without having time to dissipate and radiate energy.
X-ray emission line profiles are often used
   as a diagnostic of space-time properties
   and conditions near the inner-disc radius
	\citep[see e.g.][]{fabian1989,stella1990,laor1991,
	fabian2000,fuerst2004,younsi2012}.
However, the ISCO location does not have a simple analytic solution
   when a massive DM envelope is present.
The gravity of the DM envelope modifies the accretion flow dynamics,
   and hence the thermodynamics and radiative properties of the inner disc.
Accretion discs around black holes
   in the presence and in the absence of a massive DM envelope
   would show different spectral profiles
	\citep[cf.][]{joshi2011,joshi2014,bambi2013}.
Thus,
   black hole parameter estimates derived without accounting for DM
   could give incorrect results.

(iv) Interferometric imaging of SMBH in nearby galaxies
   will be possible with the development of
   the Event Horizon Telescope (EHT)\footnote{{\tt http://www.eventhorizontelescope.org/}}
   and the Greenland Telescope (GLT)\footnote{{\tt http://www.cfa.harvard.edu/greenland12m/}}.
A SMBH that is heavily enveloped by DM
   might show an `event horizon' shadow smaller 
   than that expected from
   stellar-kinematic mass deductions
   \citep[cf.][]{falcke2000,nusser2004,doeleman2008}.

\subsection{Dark matter physics and microphysics}
\label{s.darkmatter}

There are many theories of DM physics
   that can viably describe the gravitational fields of galaxy haloes.
At galaxian scales,
   the only essential requirements are that the unknown material
   is electromagnetically invisible
   and has no discernable effect on nucleosynthesis
   or the stability of normal stars.
Since halo shapes are spheroidal,
   the DM seems unable to lose energy
   as readily as the radiatively cooling gas in classic astrophysical discs
   (although see \citealt{fan2013a}).

Often DM is simply assumed to be collisionless:
   practically an invisible self-gravitating dust.
This provides an easy prescription for cosmological simulations
   employing $N$-body methods.
However, observations do not confirm
   the predicted density cusps
   (see Section~\ref{s.introduction}).
Simulations also over-predict numbers of dwarf galaxies
	\citep{klypin1999,moore1999a,donghia2004,tikhonov2009,
		zwann2010,klypin2014},
   and dense large satellites that are unseen in reality
	\citep{boylan2011,boylan2012,miller2013,
		tollerud2014,kirby2014,kimmel2014}.
The question then is:
   what variety of modification or alternative theory is necessary?
Suppose that some process drives 
   the DM phase-space distribution function to become locally isotropic
   and proportional to a power of the single-particle energy,
   $f\propto(-E)^{(F-3)/2}$.
Then a polytropic relation (\ref{eq.state})
   emerges 
	\citep{camm1952}.
In collisionless DM simulations,
   the cuspy haloes have
   $Q$ following a power of $r$
   when assuming $F=3$
	\citep{taylor2001,ludlow2011},
   which implies a constant-$Q$ singular polytrope
   for some non-integer $F$ value.
For those models,
   the severest challenge is to explain why
   kpc cores occur in real haloes.
Driving processes might involve
   shaking by an elaborate baryonic feedback \citep[e.g.][]{peirani2008a},
   or collective phenomena similar to bar-mode instabilities.
This needs fine-tuning to achieve a realistic core radius.
When simulations invoke ad~hoc feedback recipes,
   these can be decisive or ineffectual,
   depending on numerical implementation
	\citep[e.g.][]{governato2010,vogelsberger2014s}.

The polytropic condition is claimed to be
   a natural equilibrium for self-gravitating systems,
   according to the \cite{tsallis1988} conjecture of extended thermostatistics.
Collisionless spheres
   may settle as `stellar polytropes'
   \citep{plastino1993,vignat2011}.
Our parameter $F$
   is linearly related to Tsallis' extensivity parameter,
   which is a non-integer.
\cite{feron2008}
   find that stellar polytropes are a poor representation
   of cuspy haloes that emerge in numerical simulations.
However, this is not a fatal criticism of the thermostatistical models,
   since real observed galaxies have cored (not cuspy) profiles.

Another possibility is that
   DM is adiabatic and self-interacting (SIDM).
The polytropic equation of state (\ref{eq.state})
   arises from basic thermodynamics,
   in the absence of complications such as phase changes.
SIDM interactions may consist of direct interparticle scattering,
   short-ranged Yukawa interactions
   or long-ranged dark forces analogous to magnetism
	\citep[e.g.][]{spergel2000,ahn2005,buckley2010,ackerman2009,loeb2011}.
If the fluid consists of point-like particles
   with only translational motions then $F=3$.
This is a common, unquestioned assumption,
   algorithmically built into many simulations of weakly interacting SIDM
	\citep{moore2000,yoshida2000b,dave2001,
	vogelsberger2012,rocha2013,peter2012,vogelsberger2014s}.
However, if DM has additional internal energy then $F>3$,
   e.g. ${\frac12}kT$ for each degree of freedom
   of rotational kinetic energy of `dark molecules'.
For diatomic dark molecules, $F=5$.
The $F$ value increases if DM particles have more composite complexity.
Independently, some efforts to reconcile
   direct detection experiments
   invoke composite or inelastic DM
	\citep[e.g.][]{smith2001,chang2009,alves2010,kaplan2010a,kaplan2010b}.
The astroparticle physics implications are increasingly recognised
	\citep[e.g.][]{cline2013b,cline2014,boddy2014}.
This possibility is inherently beyond the scope of $N$-body codes.
The quantity $F$ might effectively vary
   in some DM theories
   that yield pressure anisotropies and/or
   a more complicated equation of state
   \citep[e.g.][]{sobouti2009,harko2011b,harko2012}.
For now, we assume constant $F$.

The nature of SIDM is still under debate. 
It is possible that SIDM is not a gas
   but a scalar field or boson condensate
	\citep[e.g.][]{sin1994,ji1994,lee1996,hu2000}. 
A polytropic equation of state can be obtained from some boson models, 
  with the value of $F$ depending on the self-coupling potential in the lagrangian.
Many works assume $F=2$
   with $s$ and $Q$ fixed universally by particle properties
	\citep[e.g.][]{goodman2000,arbey2003,boehmer2007,harko2011c,chavanis2011}, 
   but other $F$ values are possible 
	\citep{peebles2000}.  
It was also suggested that phase changes can occur in bosonic DM 
   and this would alter the spatial variations in the properties 
   of large astrophysical objects  
	\citep[see e.g.][]{arbey2006,slepian2012}. 

Alternatively, 
   DM may consist of neutral fermions
   \citep{dodelson1994}.
Warm DM made of sterile neutrinos ($\sim1$--$7$keV mass range, 
   depending on the primordial particle distribution) might 
   decay, producing X-ray emission lines  
	\citep{boyarsky2014,bulbul2014}.
In this case, 
   a degenerate phase can act as a cored polytrope with $F=3$
   \citep[e.g.][]{munyaneza2005,munyaneza2006,richter2006,chan2008,destri2013}.
For fermionic DM,
   the Pauli exclusion principle implies a universal maximum phase-space density,
   $Q\le Q_\mathrm{max}$.
In a DM halo,
   $q\le Q_\mathrm{max}c^F(8\pi GR^2/3c^2)\chi^{(F-2)/2}$.
Assuming that the haloes have roughly the same mean density
   gives $q\le Q_\mathrm{max}(c^F/\rho_\mathrm{v})\chi^{F/2}$. 
Either way, a region of the $(\chi,q)$ map is excluded above a diagonal line.
For sufficiently low $\chi$, non-singular solutions are excluded.
This implies that isolated haloes cannot form below some critical mass
   (if nonsingular)
   or otherwise they must have a singular, terraced or BH-dominated centre.
Assuming that dwarf galaxies obey this limit,
   observationally inferred $Q$ values
   reveal or exclude the candidate particle properties
	\citep[e.g.][]{tremaine1979,boyarsky2009,destri2013,
	devega2014,horiuchi2014,domcke2014}.
Those studies implicitly assume pointlike particles ($F=3$).
The wider possibilities of $F\neq3$ fermions remain unchecked.

Whatever the fundamental nature of DM,
   its distribution must deform
   within the gravitational sphere of influence of a SMBH.
A dark density spike emerges either as a static equilibrium,
   or as the result of gradual capture of DM at the horizon
\citep{ipser1987,quinlan1995,gondolo1999,
	ullio2001,macmillan2002,peirani2008b}.
A stellar density cusp may also develop in this region
   \citep{bahcall1976,young1980}.
In dense galaxy nuclei,
   the gravitational scattering of DM by the stars
   renders the halo {\em indirectly} collisional
   \citep{ilyin2004,gnedin2004a,zelnikov2005b,vasiliev2008,
	merritt2004,merritt2010}.
Unless DM annihilation or other effects overrule the dynamics,
   a polytropic description applies in the stellar cusp.
For standard DM with point-like particles ($F=3$)
   the dark spike profile is $\rho\sim r^{-3/2}$.
If there is internal energy then $F>3$
   and the spike is steeper,
   $\rho\sim r^{-F/2}$.
If $F>6$ then the mean-free-path of self-scattering
   ($\lambda\propto \sigma^4/\rho\sim r^{(4-F)/2}$)
   shortens and vanishes at small radii in the spike,
   which justifies an adiabatic SIDM treatment
   regardless of DM collisionality properties in the outer halo.
Our formulae (\ref{eq.m.hole}) and (\ref{eq.m.value})
   relating the black hole mass ($m_\bullet$)
   to the DM properties
   therefore should hold locally in galactic nuclei.

The observational evidence that galaxies possess kpc-sized dark cores
   is well modelled by polytropic density profiles.
Relative to the global $M$ and $R$,
   core sizes tend to shrink as $F$ increases.
Scaling relations among disc galaxies imply high $F$
   \citep{nunez2006,zavala2006}.
From the kinematics of elliptical galaxies,
   the inference is $7\la F\la9$
   \citep{saxton2010}.
Models of galaxy clusters comprising DM
   and cooling gas inflows
   predict realistic core sizes
   and enable realistic $m_\bullet$
   if $7\la F<10$
   \citep{saxton2008,saxton2014a}.
Now our simple analysis of BH plus adiabatic haloes
   also supports this range of $F$.
The case of $F=9$ naturally leads to a rule $m_\bullet\sim\sigma^{4.5}$.
If velocity dispersions of DM and stars
   both follow the shared gravitational potential,
   then this matches the observed correlation,
   $m_\bullet\sim\sigma_\bigstar^{4.5}$.

\subsection{Globular clusters as a tracer}

Globular clusters (GC) inhabit the host galaxy's halo
   and provide a useful physical probes where other visible tracers are rare.
The GC swarm diminishes with distance from the core,
   but can also develop central deficits
	\citep{capuzzo2009}.
GC consist of uniformly old and metal-poor stellar populations.
They appear to lack DM of their own:
   stellar mass suffices to explain the internal kinematics
   \citep[e.g.][]{heggie1996,baumgardt2009,sollima2009,lane2010,conroy2011,
		hankey2011,bradford2011,sollima2012,ibata2013}.

GC formation was either a purely baryonic process,
   or else their miniature DM haloes were ablated later.
The oldest GC apparently formed in brief single starbursts 
   comparable to a dynamical time of the proto-galaxy,
   perhaps caused by thermal instabilities or shock compressions
   of clouds in the halo
   \citep[e.g.][]{searle1978,fall1985}.
Newer (metal rich) GC may form from shocked gas in wet mergers
   \citep{ashman1992,zepf1993,whitmore1995,hancock2009,whitmore2010,smith2014}.
Dry mergers of galaxies
   combine preexisting GC swarms
   and preserve the ratios of SMBH, stellar and GC masses.
GC on radial orbits traversing the inner galaxy
   can be destroyed by tidal shocking
   \citep[e.g.][]{ostriker1972,fall1977,gnedin1997,gnedin1999,fall2001}.
Compared to ellipticals,
   disc galaxies seem more efficient as GC destroyers
   or less efficient GC formers.
	\citep[e.g.][]{harris1988,georgiev2010}.
The surviving GC population depends on:
   the primordial baryonic mass endowment;
   the subsequent formation and destruction processes;
   and the bredth and depth of the halo potential binding GC to the galaxy.
By the virial theorem or Jeans modelling,
   the radial velocity dispersion of the GC system
   is proportional to the depth of the halo potential.

Since the GC swarm traces aspects and properties of the whole galaxy halo,
   it is significant that GC observables
   correlate with the SMBH ($m_\bullet$)
	\citep{spitler2009,burkert2010,harris2011,harris2013}.
\cite{snyder2011}
   interpret the SMBH-GC correlations
   as consequences of the depth
   of the galaxy bulge's gravitational potential.
\cite{sadoun2012}
   relate the velocity dispersion of the GC system,
   $m_\bullet\sim\sigma_\mathrm{gc}^\beta$
   with $\beta=3.78\pm0.53$.
\cite{pota2013b}
   also linked $m_\bullet$ with $\sigma_\mathrm{gc}$
   ($3\la\beta\la6$ or $\beta\approx4.45$ on average);
   and \cite{rhode2012}
   found $\beta\approx5.3$ or $5.9$.
These $\sigma_\mathrm{gc}$ relations have great implications.
This correlation could be evidence
   of a link between SMBH formation
   and the halo properties,
   not merely the properties of the stellar bulge.
The stronger the $m_\bullet$--$\sigma_{\rm gc}$ relation is,
   the less likely that these components are controlled by BH feedback,
   and the more likely that it depends somehow
   on the underlying DM potential.

\cite{burkert2010} and \cite{rhode2012}
   have a different interpretation:
   attributing the correlation to the effect of mergers later on
   (more mergers produce more GCs and a bigger SMBH).
We suggest that the correlation would not be so tight
   if the individual merging blocks did not already
   have a correlation on their own.
Furthermore, mergers cannot have been the controlling process
   in bulgeless thin-disc galaxies
   that host a SMBH but have never experienced a major merger
   (Section~\ref{s.ltg}).
Mergers cannot be the universal explanation.
Instead we propose that the halo controls the SMBH origin
   and the GC properties separately.
In each large galaxy,
   there will be a fraction of large GCs produced {\em in situ}
   during the initial collapse,
   and a fraction coming later
   from the disruption of nucleated satellite galaxies. 
Stellar populations and orbital kinematics are usually
   clues to which is which 
   (e.g. M54 and $\omega$~Cen may be satellite accretions).
It would be interesting to
   predict the implications if local GCs 
   are those formed without a DM potential well, 
   and those coming from accreted
   galaxies are formed at the bottom of that galaxy's DM potential, 
   perhaps even with their own nuclear black holes.

\subsection{SMBH formation and accretion}

Our equilibrium configurations
   do not distinguish how the central object originated.
We simply have a non-evolutionary description of the endpoint
   after the inner halo attains approximate pressure balance.
Our model $m_\bullet$ limits do not apply
   while a system is dynamically disturbed, asymmetric
   and evolving into another state.
However 
   the most realistic equilibrium solutions tend to have small $\psi$ values,
   meaning that a dark envelope is a significant presence around the horizon.
This suggests that DM accretion may be relevant to
   SMBH seeding and growth.
We are aware of at least three scenarios.
Steady growth is possible via \cite{bondi1952} accretion of fluid
   \citep[e.g.][]{munyaneza2005,munyaneza2006,richter2006,
	peirani2008b,guzman2011b,guzman2011a,pepe2012,lorac2014}
   or gradual capture of collisionless orbiting particles
   accompanied by loss-cone refilling.
   \citep[e.g.][]{peebles1972b,ullio2001,vasiliev2008}.
If the dark matter self-interactions are weak
   (with a kpc-sized mean-free-path)
   but heat conduction is significant
   then gravothermal instability could form a SMBH
   \citep{ostriker2000,hennawi2002,balberg2002a,balberg2002b}.
If SIDM is a fluid with $F>6$
   then collapse may proceed via a localised gravitational instability
   in a discrete `dark gulp' 
   lasting a dynamical timescale of the nucleus
   \citep{saxton2008,saxton2014a}.
The gulped dark mass
   could be an appreciable fraction of the SMBH total.

Initiating this process may require a steep central density gradient.
BH seeding is probably helped if there is already a steep spike
   of stars or accumulation of inflowing gas.
It may be necessary
   for baryons to become denser than some threshold,
   in order to pinch the DM
   \citep[via adiabatic contraction,][]{blumenthal1986}
   and enable collapse of the innermost DM.
Perhaps this pinching can partly explain
   the observed correlations between SMBH
   and the Sersic index of the stellar surface brightness profile
	\citep{graham2001,graham2007,savorgnan2013}.
Evaluating the collapse thresholds needs
   multi-component stability analyses,
   like \cite{saxton2013} but with a density spike.

Some comparisons of the mass function of the local SMBH population 
with the AGN and quasar luminosity distribution   
were consistent with most of the current SMBH mass 
coming from radiatively efficient gas accretion
\citep{soltan1982,salucci1999,yu2002,shankar2004,shankar2009}.
This does not invalidate our proposed scenario.
If these audits of light and mass are complete, 
they are still consistent with an initial relation 
between bulge mass and seed BH mass, in which the latter 
could have been $\la10^{-4}$ bulge mass and much less than the final SMBH mass.
That situation corresponds to $\chi\la10^{-6}$ in the $F\ga8.5$ halo models.
Spatially, $R \sim 10^{12} r_\bullet$
   would be a plausible size for a seed black hole,
   as $r_\bullet\la10^{11}$~cm
   for large galaxies with $R \sim 10$s of kpc
   ($\sim10^{23}$~cm).
These seeds could have condensed according to our predicted scaling index,
    $m_\bullet\sim\bar{\sigma}^{F/2}$,
   and then grown through \cite{eddington1918b} limited
   luminous accretion of gas.
The final observed black-hole mass
   would be $10^n$ times the seed mass,
   after $\approx n$ Salpeter timescales.
The scaling relations would rise in normalisation
   but retain the original slope:
   $m^\prime_\bullet=10^nm_\bullet\sim \sigma^{4.5}$
   (if $F\approx9$).
The index of SMBH scaling is preserved from our simplistic gasless halo model.

Note that there are always uncertainties and complications 
in the accounting of total SMBH mass and radiative efficiency 
of their growth. 
For example, recoiling SMBHs can escape their galaxies after a merger 
	\citep[e.g.][]{redmount1989,menou2001,haiman2004,madau2004,baker2006,
		gonzalez2007,campanelli2007a,campanelli2007b,schnittman2007,
		lousto2011,lousto2013}, 
and end up dormant in intergalactic space: in that case, simple counts 
of nuclear SMBHs would underestimate the total cosmic BH mass.
The local SMBH density may also have been 
underestimated if there is a previously unrecognized population of SMBHs 
in ultracompact dwarf galaxies \citep{seth2014}.
The presence of ultramassive BHs may require
   more accretion (via radiatively inefficient modes) than reckoned before
	\citep{mcconnell2011,mcconnell2012,vandenbosch2012,fabian2013}.
The discovery of modern-sized quasars at high redshift 
   is likely to require a faster early growth than allowed by 
   Eddington-limited luminous accretion 
	\citep{fan2004,shapiro2005,mortlock2011,venemans2013}; 
   on the other hand, the X-ray background from high-redshift AGN
   is dimmer than expected, contradicting rapid 
   radiatively-efficient gas accretion in the $z>5$ era
	\citep{willott2011,salvaterra2012,treister2013}. 
The radiative efficiency of quasar accretion may be 
lower than the standard $\eta \sim 0.1$ disk efficiency during super-critical 
gas accretion phases \citep{novak2013} or due to DM accretion; 
however, accretion can instead appear more radiatively efficient
than $\eta \sim 0.1$, for example if it taps into the BH spin 
\citep{narayan2003,igumenshchev2008,tchekhovskoy2014,lasota2014}
   or if DM envelope dominates the inner potential.
Finally, SMBHs might grow via BH-BH coalescence without 
any radiative emission; however, constraints set by the cosmic 
gravitational-wave background
   imply that steady accretion (of gas or DM) dominates
	\citep{shannon2013}. 
Thus, the issues of how early SMBHs were seeded,
   the role of DM in setting the seed mass
	\citep[e.g.][]{mack2007,dotan2011,lorac2014}
   and the DM mass contribution are far from settled.

Our halo model has some similarities to
   the supermassive star scenario
   that aims to explain the early SMBH seeding.
The proposal is that a $\ga10^5m_\odot$ polytropic sphere of gas
	\citep[e.g.][]{hoyle1963a,iben1963,fowler1964,shibata2002}
   burns and collapses to produce a seed BH
   that is born supermassive,
   thereby reducing the feeding time needed to reach observed SMBH scales
	\citep[e.g.][]{begelman2006,begelman2010,johnson2013}.
The main doubt about this scenario is that
   the gas may not collapse into a single supermassive object,
   and may instead fragment into clumps and star clusters
   because of its angular momentum.
Even if a single supermassive star were formed,
   it may not survive long enough to develop a core
   and collapse into a single black hole,
   due to mass losses in intense winds.
Our model would create MBH seeds from polytropic DM instead.
Eddington limits and winds do not apply to SIDM seeding.
Whichever way real SMBH originated,
   we expect a scaling like $m_\bullet\sim\sigma^{F/2}$
   to emerge from the direct or indirect coupling
   of the SMBH and halo in equilibrium,
   since the equilibrium state
   is independent of what fed the SMBH previously.

\subsection{Late-type galaxies}
\label{s.ltg}

It has long been a puzzle to explain why
   ellipticals, lenticulars and early-type spirals
   have a nuclear SMBH,
   while many late-type spirals
   have a nuclear star cluster but no SMBH.
M33 and NGC205 are local examples of the latter
\citep{gebhardt2001,merritt2001,valluri2005}.
Even more puzzling is the fact
   that the nuclear star cluster mass versus $\sigma$ relation 
   runs parallel to the $m_\bullet$ scaling relations
	\citep{graham2009,graham2012b,ferrarese2006}.
On the other hand,
   some bulgeless galaxies do possess a nuclear SMBH
   \citep[e.g.][]{filippenko2003,peterson2005,
	shields2008,araya2012,simmons2013,reines2013}.

\cite{salucci2000b}
   observed that late-type galaxies have SMBH
   that are undersized compared to 
   the usual trend with bulge mass ($M_\bigstar$).
It is arguable those galaxies only have pseudo-bulges
   (evolved quiescently from the disc via secular processes),
   whereas SMBH correlate with classical bulges
	\citep{kormendy2011}.
Alternatively,
   perhaps the $m_\bullet$ relation
   bends downwards in the low-$M_\bigstar$ domain
	\citep{graham2012a,scott2013}
   and the SMBH relation to $\sigma$ is straighter.
This hints that $\Phi$ plays the fundamental role,
   consistent with our thesis linking the SMBH to the halo.
Either way, the hints of some dependence on luminous morphology
   (besides the DM halo)
   deserve an explanation within our theory.

It is worth noting some exemptions
   from the SMBH mass prediction of
   equations (\ref{eq.m.hole}) and Appendix~\ref{s.absolute}.
If the velocity dispersion $\sigma$
   is non-relativistic everywhere in the profile,
   then there need not be an event horizon at the centre.
A nonsingular halo does not grow any central compact mass.
This is the lowest entropy condition available.
We propose that protogalaxies condensed in this initial state,
   and some would grow quiescently
   (without major mergers or gas expulsions)
   till the present epoch.
Those are tranquil disc galaxies,
   near the nonsingular border,
   lacking classical bulges,
   and having undersized SMBH or none at all.
For other galaxies,
   tidal harassment or minor mergers would raise the entropy
   (lowering $q$),
   inducing a more centrally peaked density profile.
Perhaps if the central DM becomes concentrated enough,
   a seed BH forms.
Subsequent large-scale gas inflows accrete onto the SMBH in a quasar phase.
These galaxy haloes enter the `plateau region';
   they follow the maximum $m_\bullet/M$ scaling relation.
For those that suffer more major mergers,
   the luminous disc converts partially into a classical bulge,
   or totally into an elliptical.
In contrast,
   for the undisturbed, high-$q$ nonsingular galaxies,
   if the inner halo never became dense enough,
   it does not form the initial black hole,
   and the same large-scale gas inflows produce a nuclear star cluster.
The mass in this nuclear star cluster
   is comparable to the baryonic mass
   that would have fed the SMBH.
We speculate that the knee in the $M_\bigstar$ correlation
   \citep{graham2012a}
   or the underweight SMBH of late-type galaxies
   \citep{salucci2000b}
   may occur:
\begin{enumerate}
\item
because the latest-type galaxies
   are near the high-$q$ nonsingular border
   and their $m_\bullet/M$ is below the relations
   in Fig.~\ref{fig.ribbons};
   or
\item
because these galaxies are near one of the knees
   in a ribbon relation such as those in Fig.~\ref{fig.ribbons}; or
\item
the bulge is incidental and the halo determines $m_\bullet$.
\end{enumerate}

Though it is beyond the scope of our spherical modelling,
    we speculate that the angular momentum of the halo
    and gas may also affect the outcome.
If the inner halo possesses too much angular momentum
    (and cannot shed it via large-scale dark turbulence)
    then rotational support inhibits collapse.
If the baryons have effective rotational support,
    then they may not achieve the central densities
    needed to trigger the inner halo to condense a seed.
The result is a pure disc galaxy without a central black hole.

\subsection{Stellar components}

Our gasless and starless model is a simplification.
In principle, a galaxy's stellar mass distribution
   affects the SMBH/halo equilibrium to some extent.
In galaxy clusters,
   \cite{saxton2008,saxton2014a}
   found that the continuity requirements of gas inflows
   impose {\em lower} limits on $m_\bullet$,
   however inserting a central galaxy's stellar profile
   did not alter these constraints greatly.
An isolated elliptical galaxy's stellar spheroid
   compresses the dark core slightly
	\citep{saxton2010,saxton2013}.
Nonetheless
   DM always dominates in the outer halo.
DM should also dominate baryons at the centre:
   within the innermost stellar orbit,
   and perhaps throughout the SMBH sphere of influence.
Visible matter is most influential at medium radii
   (kpc for an elliptical galaxy).

Our present models omit stellar profiles,
   as we are most interested in the link
   between the DM halo and the SMBH.
Because observations already show that these properties correlate,
   we suspect that the stellar mass
   does not dominate SMBH scaling relations outright.
This motivates our comprehensive exploration of baryon-free configurations.
Our model has two components and three key parameters:
   thermal degrees of freedom ($F$),
   compactness ($\chi$)
   and entropy ($s$, via $Q$ and $F$).
Adding one more density component
   will increase the complexity of the formulation,
   if we want a self-consistent treatment.
This topic is worth a separate study,
   and we intend to resume it elsewhere.
However, we would also like to comment qualitatively here.
The addition of a stellar spheroid entails
   three more free variables:
   total stellar mass ($M_\bigstar$),
   a half-light radius ($R_\mathrm{e}$)
   and Sersic shape index ($n$),
   vastly increasing the system's dimensionality.
We ran restricted tests of $F=9$ models
   where the stars comprise $10\%$ of the mass.
In a preliminary way, we note:
\begin{enumerate}
\item
If the stellar component is compact
   ($R_\mathrm{e}\ll R$)
   it exerts little influence on the scaling relations.
This is understandable since
   this bulge behaves somewhat like a central concentrated point,
   which is effectively the same as the SMBH.
\item
For any terraced or singular model,
   the DM dominates at a sufficiently small radius.
The stellar component is also sub-dominant
   at the radius of halo core and outskirts beyond $r\gg$ kpc.
The stellar potential only perturbs the DM density profile
   locally at intermediate radii.
\item
Theoretically, the worst scenario is when
   the DM and the stellar component have similar compactness.
Even then,
   we find that the basic conclusion holds,
   except that $\psi$ and $m_\bullet/M$ values
   shift across the parameter plane.
This shift is only significant near the nonsingular border.
We will leave the detailed discussion for our next paper.
\end{enumerate}
The robustness of SMBH vs halo scaling relations,
   in spite of a stellar contribution and medium radii,
   might be foreseeable on qualitative grounds.
The halo core depends on heat capacity and entropy.
The location of the event horizon
   (which sets $m_\bullet$)
   coincides with an effectively universal maximum of $\sigma^2$.
Both defining structures depend straightfowardly
   on the gravitational potential and velocity dispersion,
   which are linearly related.
Their correlation arises naturally.
Essentially and generally,
   the empirical SMBH scaling relations
   reveal how density $\rho$ is stratified
   with respect to potential $\Phi$ in galaxies.


\section{Conclusions}

We investigate the properties 
  of spherical, adiabatic self-gravitating systems  
  with the DM microphysics prescribed by an equation of state.   
These systems form a halo of DM and a central compact object.   
We have found that the halo profile is determined by two necessary parameters. 
One possible combination is the gravitational compactness $\chi$
   (equation~\ref{eq.compactness})
   and a measure of (pseudo-)entropy $s$ 
   (or equivalently the phase-space density $Q$).
Characterisation of such halo profile in terms of a single parameter 
   \citep[e.g. asymptotic or peak circular velocity: see][]
  {ferrarese2002,baes2003,zasov2005,kormendy2011,volonteri2011}  
   is therefore incomplete 
   ---  the configuration-space encompasses a variety of density profiles
   that are not merely rescaled versions of a standard profile.
The halo can be nonsingular or singular.  
Nonsingular haloes lack a SMBH,
   and they correspond to the lowest entropy condition. 
Singular haloes, which have a SMBH,
   could have one or several concentric DM cores,  
   over particular radial ranges.
The most extreme singular haloes are dominated by a central black hole, 
   together with a diffuse atmosphere of negligible mass.
When the models are projected in terms of
   the compactness of the kpc-scale DM core,
   the configuration space reduces, 
   so that the haloes almost resemble 
   the one-parameter models that are common in astrophysical practice.
Where we include nonsingular and nearly nonsingular galaxies
   besides the singular `plateau' cases,
   the ribbon-like relations
   become upper limits on $m_\bullet/M_\mathrm{c}$.

The SMBH mass scales with the characteristic velocity dispersion,
   $m_\bullet\sim\sigma^{F/2}$, 
   with effective thermal degrees of freedom $F$ as the scaling index. 
Given that bulge stars and DM particles
   bound in the same potential well
   have similar velocity dispersions,
   the observed $m_\bullet$ vs $\sigma_\bigstar$ scaling relation
   indicates that $F\ga7$ for the dark halo.
The recently observed correlations between SMBH
   and velocities of halo GC swarm ($\sigma_\mathrm{gc}$)
   are also consistent with this conclusion.
The consistency of these correlations
   (especially GC properties at the far outskirts)
   supports an idea that SMBH scaling relations
   are controlled by the underlying DM potential
   rather than by AGN feedback (which operates at the centre).
The finding that $F\ga 7$
   implies that DM has large effective degrees of freedom,
   which we might interpret as a large heat capacity,
   or perhaps a steep index of a self-interaction potential.
These values agree with the range indicated in some previous modelling
   of elliptical galaxies and galaxy clusters
	\citep{saxton2008,saxton2014a,saxton2010}.
   
These models also tend to predict
   that a dense dark envelope surrounds the SMBH.
In at least some systems,
   the envelope may have non-negligible density
   compared to the SMBH itself.
In extreme cases, the dark envelope outweighs the SMBH.
This might have observable consequences
   in the relativistic vicinity of the event horizon.
Useful tests might involve
   apparent sizes of SMBH horizons,
   the tidal disruption of stars,
   and the inner structure of AGN accretion discs.



\section*{Acknowledgments}

We thank the referee, P.~Salucci,
   for helpful criticisms and suggestions
   that improved the scope and focus of our results and commentary.
We thank
A.W.~Graham for discussions of SMBH scaling, 
and M.~Cropper for discussions of GC populations.
This work has made use of NASA's Astrophysics Data System.
Our calculations employed mathematical routines from
the {\sc Gnu Scientific Library}.
This publication has made use of code written by
   James R. A. Davenport.\footnote{%
	\tt{http://www.astro.washington.edu/users/jrad/idl.html}
	}
Specifically, the figures' colour
   scheme\footnote{%
	\tt{http://www.mrao.cam.ac.uk/{\textasciitilde}dag/CUBEHELIX/}
	}
   was developed by \cite{green2011}.

\bibliographystyle{mn2e}
\bibliography{jour,blob}

\appendix

\section{SMBH prediction in absolute terms}
\label{s.absolute}

The equation (\ref{eq.m.hole})
   for the SMBH mass can be written in various absolute units
   for practical applications.
The choice of units depends on context.
For example,
   in the vicinity of the dark envelope and the circumnuclear region,
   velocity dispersions are almost relativistic.
DM densities could become comparable to that of baryonic matter on Earth.
In units suiting that environment,
   the SMBH mass in solar units is
\begin{equation}
	{{m_\bullet}\over{m_\odot}}
	\approx
	{{4.2919\times10^9}\over{\sqrt{\eta^3\psi}}}
	\left({{F+2}\over{1-\chi}}\right)^{\frac{F}{4}}
	\left({{1\,\mathrm{kg}\,\mathrm{m}^{-3}}\over{\rho}}\right)^{\frac{1}{2}}
	\left({{\sigma}\over{c}}\right)^{\frac{F}{2}}
	\ .
\end{equation}

Farther out, in the kpc-scale core of the galaxy's halo,
   typical velocities drop to the order of
   $100~\mathrm{km}~\mathrm{s}^{-1}$.
DM core densities are multiples or fractions of
   $1~m_\odot~{\rm pc}^{-3}$.
In these terms,
   the predicted central mass (solar units) is
\begin{eqnarray}
	{{m_\bullet}\over{m_\odot}}
	&\hspace{-3mm}\approx&\hspace{-3mm}
	{{1.6495\times10^{19}}\over{\sqrt{\eta^3\psi}}}
	0.018264^F
	\left({{F+2}\over{1-\chi}}\right)^{\frac{F}{4}}
	\nonumber\\&&
	\times\left({
		{{1\,m_\odot~{\rm pc}^{-3}}\over{\rho}}
	}\right)^{\frac12}
	\left({
		{{\sigma}\over{100\,{\rm km}~{\rm s}^{-1}}}
	}\right)^{\frac{F}{2}}
	\ .
\end{eqnarray}
An equivalent logarithmic form says
\begin{eqnarray}
	\log_{10}\left({
		{{m_\bullet}\over{m_\odot}}
	}\right)
	&\hspace{-3mm}\approx&\hspace{-3mm}
	19.217
	-1.7384\,F
	+{F\over4}\log_{10}\left({
		{ {F+2}\over{1-\chi} }
	}\right)
	\nonumber\\&&
	-{1\over2}\log_{10}\left({
		\eta^3\psi
	}\right)
	-{1\over2}\log_{10}\left({
		{{\rho}\over{1\,m_\odot\,\mathrm{pc}^{-3}}}
	}\right)
	\nonumber\\&&
	+{F\over2}\log_{10}\left({
		{{\sigma}\over{100\,\mathrm{km}~\mathrm{s}^{-1}}}
	}\right)
	\ .
\label{eq.m.value}
\end{eqnarray}
The third term on the right side is $<2.7$ when $\chi\ll1$.
The astronomical mass range $m_\bullet\la10^{10}m_\odot$
   implies that either $F>6$ (in the second term of the right side)
   or there is a large correction factor $\eta^3\psi$
   (in the fourth term on the right).

\section{Model homologies and scale-invariant parameterisation}
\label{s.scaling}

Given a particular polytropic halo model,
   a family of homologous models can be formed
   by multipling each quantity $y$ by a scale factor $X_y$.
Since we take the speed of light as an absolute reference scale
   for velocity dispersions, escape velocities
   and gravitational potentials,
   we necessarily have $X_\sigma=1$,
   $X_V=X_m/X_r=1$ and $X_\Phi=1$.
It follows that model masses and distances
   must rescale by the same factor,
   $X_m = X_r \equiv X$,
   and densities rescale as $X_\rho=X_m/X_r^3=X^{-2}$.
The phase-space density rescales as
   $X_Q=X_\rho/X_\sigma^F = X^{-2}$.
For example, if we choose to standardise a set of models
   so that they have the same total mass $M$,
   we could transform the phase-space densities as
   $Q\rightarrow Q/R^2$.

We prefer to classify and compare models
   in terms of their dimensionless properties
   that remain constant under the homology transformations.
Dimensionless quantities such as $\chi$, $\eta$ and $\psi$
   remain constant under the homology transformations.
If the outer boundary conditions are known,
   then it is possible to define a dimensionless variable related to $Q$,
   for instance $q\equiv QV^F/\bar{\rho}$ for which $X_q=1$.
Similarly, $l\equiv M^2Q$ for which $X_q=X_l=1$.
The properties of the central object
   are best described in terms of invariant fractional quantities
   such as the $m_\bullet/M$ and $r_\bullet/R$.

\section{Entropy calculation}

The \cite{bekenstein1973} entropy of an event horizon is
   $S_\bullet=kA/4l_\mathrm{P})^2$
   where $A$ is the surface area,
   $k$ is Boltzmann's constant,
   $l_\mathrm{P}=Gm_\mathrm{P}/c^2$ is the Planck length,
   and $m_\mathrm{P}$ is the Planck mass.
Substituting the area of the inner boundary of our model,
   $A\approx4\pi r_\bullet$,
   we have
   $S_\bullet=\pi k(r_\bullet/l_\mathrm{P})^2$,
   which simplifies:
\begin{equation}
	S_\bullet
	=\pi k\left({{c^2r_\bullet}\over{Gm_\mathrm{P}}}\right)^2
	=\pi k\left({{c^22Gm_\bullet\eta}\over{Gm_\mathrm{P}c^2}}\right)^2
	=4\pi k\left({{m_\bullet\eta}\over{m_\mathrm{P}}}\right)^2
	\ .
\end{equation}

Since the total mass of the system is $M$,
   the mass of DM outside the SMBH is $M-m_\bullet$,
   and the number of dark particles is $N=(M-m_\bullet)/\mu$.
The DM halo entropy is
   $S_\mathrm{d}=-Nk\ln({Q/Q_0})$.
For the total entropy,
\begin{eqnarray}
	S
	&\hspace{-2mm}=&\hspace{-2mm}
	4\pi k\left({
		{M\over{m_\mathrm{P}}}
	}\right)^2
	\left({
		{{m_\bullet}\over{M}}
	}\right)^2\eta^2
	-{M\over\mu}\left({
		1-{{m_\bullet}\over{M}}
	}\right)\,\ln\left({{Q}\over{Q_0}}\right)
\nonumber\\
	&\hspace{-2mm}=&\hspace{-2mm}
	{{Mk}\over{\mu}}\left[{
		4\pi{{M\mu}\over{m_\mathrm{P}^2}}
		\left({ {m_\bullet}\over{M} }\right)^2\eta^2
		-\left({
			1-{{m_\bullet}\over{M}}
		}\right)\,\ln\left({{Q}\over{Q_0}}\right)
	}\right]
	\ .
\label{eq.all.entropy}
\end{eqnarray}
The first term (entropy of the horizon) dominates if
   $\mu\gg m_\mathrm{P}^2/M$,
   and the second term (entropy of the DM halo) dominates if
   $\mu\ll m_\mathrm{P}^2/M$.
Also note the trivial algebraic identity,
\begin{equation}
	{{m_\bullet}\over{M}}\eta
	={{m_\bullet r_\bullet}\over{M r_\mathrm{s}}}
	={{c^2m_\bullet r_\bullet}\over{2GM m_\bullet}}
	={{c^2r_\bullet}\over{2GM}}
	={{r_\bullet}\over{R\chi}}
	\ .
\end{equation}
For fixed $\chi$, the ratio $m_\bullet/M$ is a monotonic function of $Q$,
   and $\eta$ remains on the order of $1$.
Also for fixed $\chi$, the ratio $r_\bullet/R$ is monotonic in $Q$
   except for wrinkles within one dex of the non-singular border.
Therefore, if the the right term of (\ref{eq.all.entropy}) dominates
   then $S$ is monotonic in $Q$;
   and if the left term dominates then $S$ is also monotonic in $Q$
   (except for subtle features near the nonsingular boundary).

\label{lastpage}
\end{document}

%% file: aas_macros.tex
\def\jref@jnl#1{{\rm#1}}

\def\aj{\jref@jnl{AJ}}                   
\def\araa{\jref@jnl{ARA\&A}}             
\def\apj{\jref@jnl{ApJ}}                 
\def\apjl{\jref@jnl{ApJ}}                
\def\apjs{\jref@jnl{ApJS}}               
\def\ao{\jref@jnl{Appl.~Opt.}}           
\def\apss{\jref@jnl{Ap\&SS}}             
\def\aap{\jref@jnl{A\&A}}                
\def\aapr{\jref@jnl{A\&A~Rev.}}          
\def\aaps{\jref@jnl{A\&AS}}              
\def\azh{\jref@jnl{AZh}}                 
\def\baas{\jref@jnl{BAAS}}               
\def\jrasc{\jref@jnl{JRASC}}             
\def\memras{\jref@jnl{MmRAS}}            
\def\mnras{\jref@jnl{MNRAS}}             
\def\pra{\jref@jnl{Phys.~Rev.~A}}        
\def\prb{\jref@jnl{Phys.~Rev.~B}}        
\def\prc{\jref@jnl{Phys.~Rev.~C}}        
\def\prd{\jref@jnl{Phys.~Rev.~D}}        
\def\pre{\jref@jnl{Phys.~Rev.~E}}        
\def\prl{\jref@jnl{Phys.~Rev.~Lett.}}    
\def\pasa{\jref@jnl{PASA}}               
\def\pasp{\jref@jnl{PASP}}               
\def\pasj{\jref@jnl{PASJ}}               
\def\qjras{\jref@jnl{QJRAS}}             
\def\skytel{\jref@jnl{S\&T}}             
\def\solphys{\jref@jnl{Sol.~Phys.}}      
\def\sovast{\jref@jnl{Soviet~Ast.}}      
\def\ssr{\jref@jnl{Space~Sci.~Rev.}}     
\def\zap{\jref@jnl{ZAp}}                 
\def\na{\jref@jnl{New~Astronomy}}        
\def\nar{\jref@jnl{NewAR}}               
\def\nat{\jref@jnl{Nature}}              
\def\iaucirc{\jref@jnl{IAU~Circ.}}       
\def\aplett{\jref@jnl{Astrophys.~Lett.}} 
\def\apspr{\jref@jnl{Astrophys.~Space~Phys.~Res.}}
\def\bain{\jref@jnl{Bull.~Astron.~Inst.~Netherlands}} 
\def\fcp{\jref@jnl{Fund.~Cosmic~Phys.}}  
\def\gca{\jref@jnl{Geochim.~Cosmochim.~Acta}}   
\def\grl{\jref@jnl{Geophys.~Res.~Lett.}} 
\def\jcap{\jref@jnl{JCAP}}      %
\def\jcp{\jref@jnl{J.~Chem.~Phys.}}      
\def\jgr{\jref@jnl{J.~Geophys.~Res.}}    
\def\jqsrt{\jref@jnl{J.~Quant.~Spec.~Radiat.~Transf.}}
\def\memsai{\jref@jnl{Mem.~Soc.~Astron.~Italiana}}
\def\nphysa{\jref@jnl{Nucl.~Phys.~A}}   
\def\physrep{\jref@jnl{Phys.~Rep.}}   
\def\physscr{\jref@jnl{Phys.~Scr}}   
\def\planss{\jref@jnl{Planet.~Space~Sci.}}   
\def\procspie{\jref@jnl{Proc.~SPIE}}   